\documentclass{emulateapj}

\usepackage{graphicx, subfigure}
\usepackage{amssymb}
\usepackage{amsmath}
\usepackage{epstopdf}
 \usepackage{hyperref}
\usepackage{color}
\usepackage{placeins}
\usepackage{natbib}
\usepackage{enumerate}
\usepackage{textcomp}
\usepackage{footmisc}
\usepackage{xspace}

\newcommand{\bea}{\begin{eqnarray}}
\newcommand{\eea}{\end{eqnarray}}
\def\beq{\begin{equation}}
\def\eeq{\end{equation}}

\def\r{{\it RHESSI}\ }
\def\f{{\it Fermi}\ }

\def\stereo{{\it STEREO}\ }
\def\konus{{\it Konus}-WIND\ }

\def\t{{\theta}}

\def\a{{\alpha}}
\def\b{{\beta}}
\def\D{{\Delta}}
\def\d{{\delta}}
\def\g{{\gamma}}
\def\e{\epsilon}
\def\G{\Gamma}

\def\he3{$^3$He\,}
\def\he4{$^4$He\,}

\def\tsc{\tau_{\rm sc}\,}

\def\tcross{\tau_{\rm cross}\,}
\def\tesc{T_{\rm esc}\,}

\def\tloss{\tau_{\rm L}\,}

\def\nufnu{\nu f(\nu)\, }
\def\e{\epsilon\, } 
\def\tbr{\tau_{\rm brem}}

\begin{document}

\title{IMPLICATIONS OF A LOOP-TOP ORIGIN FOR MICROWAVE, HARD X-RAY, AND LOW-ENERGY GAMMA-RAY
EMISSION FROM BEHIND-THE-LIMB FLARES}
\author{Vah\'e Petrosian$^{1,2}$}
\affil{$^{1}$Department of Physics and KIPAC, Stanford University,
Stanford, CA 94305, USA\\ 
$^{2}$Department of Applied Physics, Stanford University, Stanford, CA 94305,
USA\\ 
}
\shorttitle{Loop Top emissions of Behind The Limb Solar Flares}
\shortauthors{Petrosian}

\begin{abstract}

{\it Fermi} has detected hard X-ray (HXR) and gamma-ray photons from three
flares, which according to 
\stereo occurred in active regions behind the limb of the
Sun as delineated by near Earth instruments. For two of these flares \r has
provided HXR images with sources located just above the
limb, presumably from the loop top (LT)  region of
a relatively large loop. {\it Fermi}-GBM has detected
HXRs and gamma-rays,  and RSTN has detected microwaves
emissions with similar light curves.  This paper presents
a quantitative 
analysis of these multiwavelength observations assuming that HXRs and
microwaves are produced by electrons accelerated at the LT source, with
emphasize on the importance of the proper treatment of escape
of the particles from the acceleration-source region and the trans-relativistic
nature of the analysis. The observed 
spectra are used to determine the
magnetic field  and  relativistic electron spectra. It is found that a simple
power-law in momentum
(with cut off above a few 100 MeV) agrees with all observations, but in  energy
space a
broken power law spectrum (steepening
at $\sim mc^2$) may be required. It is also shown that the production of the
$>100$ MeV photons detected by {\it Fermi}-LAT at the  LT source would
require  more energy compared to 
photospheric  emission. These
energies are smaller than that
required for electrons, so that the possibility that  all the emissions
originate in the LT cannot be ruled out on
energetic grounds. However, the differences in the light curves and emission
centroids of HXRs and $>100$ MeV gamma-rays favor a different source for the
latter.

\end{abstract}


\section{INTRODUCTION}
\label{intro}

{\it Fermi} Gamma Ray Observatory ({\it Fermi}; Atwood et al. 2009) observes the
Sun
once every other
orbit. During the past solar active phase its Large Area Telescope (LAT) has
detected $>100$ MeV photons from more than 40 solar flares. Few of these are
detected only during the impulsive phase coincident with hard X-rays (HXRs)
produced as nonthermal electron bremsstrahlung  (NTB) and nuclear gamma-ray
lines excited by
high energy ions (mostly protons) (Ackermann et al.~2012). There is considerable
evidence that the electrons are accelerated in a reconnection region near the
looptop (LT) of the flaring loops (Masuda et al. 1994; Petrosian et al.~2002;
Nitta et al.~2010; Krucker et al.~2010; Liu et al.~2013), and it is generally
assumed that this is the site of acceleration of the impulsive phase protons
(and ions) as well. But a majority of the LAT detected flares show only long
duration emission (extending up to 10's of
hours) usually rising after the impulsive phase (Ajello et al.~2014).
Some stronger flares show both impulsive and gradual emission (Ackermann et al.~
2014). Almost all LAT flares are associated with relatively fast ($>1000$ km/s)
Coronal Mass Ejections (CMEs) and are often accompanied with gradual  Solar
Energetic Particle (SEP) events.
This may indicate that the high energy particles
responsible for the LAT gamma-rays are accelerated in the CME shock environment
where the SEPs are produced. However, while SEPs are particles escaping
the upstream
region of the CME shock the gamma-ray producing particles, if originating
at the CME, most likely come from the
downstream region of the shock, with  magnetic
connection
to the higher
density solar atmosphere,  which is the only place such high
energy radiation can be produced.
This scenario has received further support from \f-LAT detection of three
flares which, as observed by {\it STEREO}, originate from active regions (ARs)
behind
the limb (BTL) of the Sun as delineated  by near-Earth instruments. The analysis
and
some preliminary interpretation of the data from \f and other instruments on
the BTL flares are presented  in Pesce-Rollins et al.(2015) and Ackermann et
al. (2017; Ack17). 

Our aim here is a more detailed modeling of the BTL flares with the 
particular focus on the determination of {\bf electron} spectra and energy
contents required
to produce the multiwavelength radiations seen in two of these flares. It should
be
noted that flares, such as these with occulted foot points, provide a
clearer view of the coronal LT source, which may be the site of particle
acceleration. Thus, the analysis presented below provides the most direct
information on the acceleration process. There are several reports of
observations of  partially occulted flares in HXRs (see, e.g. Frost \&
Dennis 1971; Krucker et al. 2007) and in gamma-ray
emissions (Vestrand \& Forrest 1993; Barat et al. 1994; Vilmer et al.
1999). More recently Effenberger et al. (2017) have provided a complete list
of \r observed  partially occulted flares combining those from cycle 24 with the
earlier list by Krucker \& Lin (2008) from cycle 23. Analysis similar to that
presented here can be carried out for any of these flare with contemporaneous 
microwave coverage.

The next section presents a summary
of the relevant observational
characteristics of these flares (all taken from Ack17). \S 3,  provides a
description of the
main focus of this paper, which is to describe the emission processes
and to determine
the characteristics of the nonthermal {\bf
electrons}
required for their production. \S 4 contains a brief discussion of the
possibility of
LT origin of $>100$ MeV gamma-rays detected by the LAT. A summary and
conclusions are presented in \S 5.

\section{REVIEW OF RELEVANT OBSERVATIONS}
\label{obs}

Multiwavelength  observations  of the BTL flares and their  analysis were
presented in Ack17, the main source of the data used here. In
Table 1 we
reproduce some  of these, and few new result from further analysis of the radio
observations, relevant for our modeling, in particular for the determination of
the broadband spectra and numbers (or energy contents) of the accelerated
particles. Only two of the three BTL flares, namely {\it SOL}2013-10-11 and {\it
SOL}2014-09-0,  had complete sets
of HXR, radio, and gamma-ray data. (For the sake of brevity,  hereafter in the
text we will refer to these as  Oct13 and Sep14 flares, respectively). For
each flare we give spectral parameters
averaged over the duration $\Delta T$ of the flare (25 and 18 min,
respectively) covering most of the impulsive phase. For HXRs we give the
$\nufnu$ energy flux.%
\footnote{This is same as $E^2dN/dE$ used in Ack17 and
$\e^2 J(\e)$ used below for HXR emissivities.} 
in units of erg cm$^{-2}$ s$^{-1}$
at 30 keV (above which the emission is dominated by NTB), the photon number
spectral index, $\gamma_X$, and a high energy
exponential cutoff
energy, $\e_{c,X}$. Most of these are obtained from \f-GBM data, which agree
with \r and
\konus data. The same parameters are also given for the LAT $>100$ MeV
gamma-rays. These are fits to the photon counts and can be used for 
modeling
these observations by either a relativistic NTB or by a pion decay  model.
In
Ack17  the photon counts were fitted directly to the
{\it thick-target} pion decay  model giving the time averaged simple power law
(accelerated) proton
indexes of 4.4 and 4.6 for these two flares, respectively. 

The radio spectral
parameters were obtained using the radio spectra shown in Figure 12 of Ack17
(also shown below; Fig.~\ref{fig:radio}).
These spectra appear to peak at a frequency $\nu_p$ falling as a  power law,
$f(\nu)\propto \nu^{-\g_r}$,  above the peak and decrease relatively steeply
below it. These  are clearly portions of optically thin and thick
gyro-synchrotron
emission with optical depth $\tau_{\nu_p} \sim 1$ at the peak.%
\footnote{The exact value depends on the index $\g_r$ and the geometry of
the source (see below).}
In Table 1  we give our
best estimates for the peak frequency $\nu_p$, $\nufnu$ flux at $\nu_p$ and
at optically thin part  $\nu=10$ GHz, and the spectral
index $\g_r$. There are many
causes of
absorption of microwave radiation in solar flares (see Ramaty \& Petrosian
1972) but the most common cause is synchrotron self absorption that gives a
spectrum $f(\nu)\propto \nu^{5/2}$ for $\nu\ll\nu_p$. The extant data is not
accurate enough to
distinguish among the various possibilities. In what follows, we will consider 
free-free and self absorptions.  We note that the microwave spectra used for
these estimation are for
the one minute interval around the peak of the light curve  where
the
particle
and
photon spectra are generally harder. This should be kept in mind when
comparing the radio with the HXR and $\g$-ray spectra that are integrated over
longer times used for HXRs.

\begin{deluxetable*}{cccccccccc}[ht]
\label{tab:SpectralData}
  \tablewidth{\textwidth}
  \tablecaption{Multiwavelength Spectral Characteristics}
  \tablehead{
  \colhead{Flux(30 keV)$^{(a)}$} &
\colhead{index $\g_X^{(b)}$} &  \colhead{$\e_{c,X}$ $^{(c)}$} & \colhead{
$\nu_p$ $^{(d)}$} & \colhead{F($\nu_p)^{(a)}$} & \colhead{Flux(10 GHz)$^{(a)}$}
&
\colhead {Index, $\g_r^{(b)}$}  & \colhead{Flux(100 MeV)}$^{(a)}$ &
\colhead{index
$\Gamma_\g^{(b)}$} & \colhead{$\e_{c,\g}^{(c)}$}
}\\
\startdata
\multicolumn{10}{c}{{\it SOL}2013-10-11 }\\
$9.0\times 10^{-8}$ & $3.2\pm 0.05$ & $\infty$ &
2.5$\pm0.3$ & $1.6\times 10^{-17}$ & $1.6\times 10^{-18}$ & $1.85\pm 0.15$ &
$4.4\times 10^{-8}$ & $-0.2\pm 0.3$ & 130$\pm$ 20\\
\hline\hline\\
 \multicolumn{10}{c}{{\it SOL}2014-09-01 }\\
$4.0\times 10^{-7}$ & $2.06\pm 0.01$ & $90\pm 7$ &
0.6 $\pm0.1$ & $1.4\times 10^{-17}$ & $1.4\times
10^{-18}$ & $0.85\pm 0.15$ & $8.2\times 10^{-7}$ & $-1.0\pm 0.3$ & 150$\pm$ 10\\
\enddata
  \tablenotetext{1}{X- and gamma-ray Flux's refer to  $\e^2J(\e)$ in  erg
cm$^{-2}$ s$^{-1}$, with $J(\e)$ as the number flux averaged over the durations
given in Table 2.  Radio fluxes $F(\nu)$ are in 
erg
cm$^{-2}$ s$^{-1}$ Hz$^{-1}$ averaged over over one minute around the peak.}
  \tablenotetext{2}{X-ray and gamma-ray indexes refer to photon numbers
$J(\e)\propto \e^{-\gamma}$; radio index  is photon energy index
$F(\nu)\propto \nu^{-\g_r}$.}
  \tablenotetext{3}{High energy cutoffs in MeV.}
  \tablenotetext{4}{Radio peak flux frequency in GHz.}
\end{deluxetable*}

In addition to the spectral observations given in Table 1,
we need a few other properties of the emission site for detailed modeling of
these
flares. Table 2 gives  some of these properties. For
each flare, we give the angular size in sr (based on \r images),
height above the photosphere (based on the  position of the AR BTL as
determined by {\it STEREO}), the distance between the centroids of \r and LAT
sources (in arc seconds), emission measure $EM$ (usually obtained from fits to
the lower
energy HXR thermal component), density as $n=\sqrt{EM/V}$ ( with $V$  the
volume of the source; see also footnote 5), the above mentioned
durations and the magnetic
field estimates based on the spectral fits to the
optically thick radio spectra as described in the next section. For the Oct13
flare the  $EM$ value obtained from \r thermal component by Fatima Da Costa
Rubio (private communication)  and
determine the volume, $V$,  from the source area times an assumed
depth
comparable to the width of the source. For Sep14 flare we do not have access to
the thermal component so we assume an upper limit for the $EM$ which gives a
lower density, appropriate for its  height above the photosphere.

\begin{deluxetable*}{cccccccc}[ht]
\label{tab:SizeData}
  \tablewidth{\textwidth}
  \tablecaption{Sizes, Density, Duration and Magnetic Field}
  \tablehead{
  \colhead{Flare} &
\colhead{Angular Size $\Omega$}&  \colhead{Height} & \colhead{D$^{(a)}$} &
\colhead{Emission Measure} & \colhead{Density} &
\colhead{Duration} &\colhead{Magnetic Field}  
}\\
\startdata
 - & sr & $10^9$ cm  &  " & cm$^{-3}$ & cm$^{-3}$ & min & G\\
 \hline\\
${\it SOL}2013-10-11$ & $1.5\times 10^{-8}$ & $1.0$ & $65$ &
$\sim 1.3\times 10^{48}$ & $\sim 1.4\times 10^{10}$ & 25 & $\sim 200-500$ 
\\
\hline\\
${\it SOL}2014-09-01$  & $4.2\times 10^{-8}$ & 20 & 275 &
 $<10^{47}$ & $< 10^{9}$ & 18 & $\sim 2-10$  \\
\enddata
 \tablenotetext{1}{Distance between the centroids od the LAT and \r sources.} 
\end{deluxetable*}

\section{MODELING OF THE LOOPTOP SOURCE}
\label{models}

We assume that particles of energy $E$ (in units $m_ec^2$) are either
accelerated outside the looptop (LT) source and  
injected  into it at a rate of ${\dot Q}(E,t)$ or they are accelerated
in this source region with a spectrum $N(E,t)$. As shown below, in either
case, because the particle energy loss time $\tloss\gg \tesc(E,t)$, the time  
spend traversing
the  LT source,
they lose a small fraction of their energy and produce {\it thin-target}
radiation. In the first case,
the
spectrum of 
particles integrated over the source region would  be 
$N(E,t)={\dot Q}(E,t)\tesc(E,t)$. If the acceleration and
emission sites are the same then ${\dot Q}(E,t)=N(E,t)/\tesc(E,t)$ will
represent
the flux of
the escaping particles. As evident the
difference between these two scenarios
is a matter of semantics,
so in what follows we will use the first scenario which gives  the number (and
energy) flux of particles that escape the LT
region (essentially at the injection  rate) to the footpoints (FPs) of the AR
located BTL,
where they lose all their energy and produce the usual
{\it thick target} FP radiations. These
emissions, the usual focus for disk flares,  are obscured by the
optically thick solar gas from near Earth instruments for a BTL flare. 
\stereo, the only satellite with a direct
view of the AR  detected EUV radiation from these flares. 

Our goal is to use the observations  to obtain the spectrum of the {\bf
injected flux}, ${\dot Q}(E)$, and {\bf accelerated particle number}, $N(E)$.
Over the small range of
energies commonly provided by observations, one can uses a simple power law 
to
describe this spectrum. However, for modeling the combined HXR and microwave
data (and the gamma-rays in case of Sep14), 
we must consider electron spectra spanning a wide range of energies;
from nonrelativistic (for production of HXRs) to extreme relativistic (for
production of radio and gamma-rays). In this case,  a
broken power law,  or a power law with an exponential cut off, could
provide a better fit.
If accelerated protons are responsible for
LAT gamma-rays we need their spectra from 300 MeV to tens of GeV, again
straddling the trans-relativistic range. This raises two important issues. 

1. When dealing with trans-relativistic spectra one must
distinguish between spectra in the energy and momentum spaces. A simple power
law  in
energy space [$N(E)\propto E^{-\d}$] will turn into a broken power law   in the
momentum space
[$N(p)\propto p^{(1-2\d)}$ at nonrelativistic  and $N(p)\propto p^{-\d}$ at
extreme relativistic momenta], and vice versa, with a break at $p\sim mc$ or
$E\sim mc^2$. This should be distinguished from the actual breaks determined by
the interplay between the parameters of the acceleration and energy transport
and loss
mechanisms. In what follows, we will consider spectra in both momentum and
energy.

2. The energy dependence of $\tesc(E)$, or the time the particles spend
in the LT
source.  The usual assumption of the  thin
target model is that particles cross the length $L$ of the source
freely with $\tesc(E,t)\sim \tcross\sim L /v$. However, this is the shortest
possible escape time. In general, $\tesc(E,t)> \tcross$  because the
source region is highly magnetized and may contain turbulence,
in which case magnetic mirroring or scattering by turbulence become important.%
\footnote{Scattering by Coulomb collisions cannot be the agent here because
then energy loss time, which is comparable to scattering time, will also be
shorter than the crossing time and we will no longer be in the thin target
regime (Petrosian \& Donaghy, 1999). This may be the case at electron 
energies $<25$ keV (see, Fig.~\ref{fig:times} below)}.
As a result the energy dependence of $\tesc(E)$  is more complex, and 
here also, it could change across the
trans-relativistic
energy. (Note that even though $\tesc(E,t)> \tcross$ the thin-target
assumption is still valid because as shown below $\tesc(E,t)<\tloss$.) In
the strong diffusion case, i.e., when scattering time $\tsc\ll
\tcross$ the accelerated 
particles random walk across the source so that $\tesc\sim \tau_{\rm
cross}^2/\tsc$
and the magnetic field variations on the scale $L$ have a small effect.
On the other hand, in the weak diffusion limit with $\tsc\gg \tcross$, and
for a converging field geometry, the
escape time is
determined by how fast particles are scatted into the loss cone, in which
case (for injected particle pitch angle distribution not highly beamed along
the field lines) $\tesc \propto \tsc$, with the proportionality constant
increasing with increasing field convergence. 
In summary we
have 

 \[ {\tesc\over \tcross}=  \left\{
 \begin{array}{lll}
      1\, & \mbox{if $\tsc\gg \tcross$, Free stream}\\ 
      \propto \tsc\, & \mbox{if $\tsc\gg \tcross$, Converging field}\\
      \tcross/\tsc\, & \mbox{if $\tsc\ll \tcross$, Strong diffusion}.
 \label{cases}
 \end{array}
      \right. \]
Combining these three cases  we obtain (see,  Malyshkin \& Kulsrud, 2001; Fig.~2
in Petrosian, 2016; [P16])
\beq
\label{tesc}
\tesc=\tcross\left(\eta+{\tcross\over \tsc}+\ln\eta {\tsc\over
\tcross}\right),
\eeq
where $\eta$ is a measure of the convergence rate of the field lines (e.g.~the
inverse of the  ratio of magnetic field at the top of the loop to where they
exit the LT source. In what follows we will use $\eta=3$ (see, e.g.~McTiernan \&
Petrosian 1991). 

In either
case, the scattering by turbulence plays a crucial role.
Relativistic
particles scatter primarily by large scale fast mode or Alfv\'en waves,
with $\tsc\propto E^{\a_{\rm er}=(2-q)}$, where $q$ is the
spectral index
of
the turbulence; for  Kolmogorov spectrum $\alpha_{\rm er}=1/3$. For
semi-relativistic and nonrelativistic particles this relation is more
complicated and does not fit a simple power law
(see Pryadko \& Petrosian, 1997 and Petrosian \& Liu, 2004). 
Chen \& Petrosian (2013; CP13), applying the
inversion method proposed by Piana et al. (2003) to  two flares,  find
energy dependences for the escape, energy loss and acceleration times,
empirically and directly from \r data, in
the
nonrelativistic regime. These results indicate that we are dealing with the
middle case in the above equation with $\alpha_{\rm nr} \sim 0.8$ and 0.2. This
is in good agreement with the distribution of $\alpha_{\rm nr}$ determined (also
empirically) based on comparison of SEP and HXR producing electron spectral
indexes (see Fig. 4 in P16). In what follows we will treat
$\alpha_{\rm nr}$ as a free parameter and use $\alpha_{\rm er}=1/3$ (or
$q=5/3$). 
\beq
\label{tsc}
\tsc(E)= \tsc_0\times (E/E_t)^{\alpha_{\rm
nr}}{1+(E/E_t)^{\a_{\rm er}}\over 1+(E/E_t)^{\a_{\rm nr}}},
\,\,{\rm with}\,\,E_t\sim 1,
\eeq

These two 
energy dependencies mold the thin target spectra.
 In Figure
\ref{fig:times} we show  the energy dependences of $\tcross, \tsc, \tesc$ (in
cyan), where we have used 
a normalization (i.e.~$\tau_{\rm sc,0}$) that 
gives them  the same
relative value with respect to crossing time and Coulomb energy losstime 
determined in CP13. Here we also give energy loss times, defined as
$\tau_L=E/{\dot E}_L$, for the  energy loss rates ${\dot E}_L$) due to 
Coulomb, bremsstrahlung, synchrotron  and 
inverse Compton (IC),%
\footnote{The IC loss rate is due to interactions  with solar
optical  photons of energy density $u_{\rm op}=L_\odot/(4\pi c r^2)$, where
$L_\odot$ is the solar luminosity and $r\sim R_\odot$ is distance from the
center of the Sun. It is identical to the synchrotron loss rate, but with an
effective magnetic field $B_{\rm eff}=\sqrt{8\pi u_{\rm op}}\sim 7$G that is
usually negligible for prevailing magnetic fields of $B>100$ G. However, for
large flaring loops, like that of Sept 14 flare,  $B_{\rm eff}>B$, and  the IC
loss becomes important.}
As evident for electron energies of $> 10$ keV of interest here the total
energy loss time (solid black) is longer than the escape time justifying the
thin-target assumption. To have a {\it thick-target} LT source we need
$\tesc\leq \tloss$. For Oct13 flare  this would require an escape time that is 
10 or 100 times
longer, for HXR and microwave ranges, respectively. This means  a 10-100
times shorter $\tsc$ or a 10-100 times higher field convergence parameter
$\eta$,
for strong and weak diffusion cases, respectively. The requirement is more
severe for Sep14 flare where the emission extends to relativistic regime so we
need $\tesc \sim 10^4$ s (i.e. $\tsc\sim 10^{-5}$ s or $\eta>10^4$, for
strong and weak diffusion cases, respectively.) 

As also evident from these
figures for most of the relevant energies we are in the weak diffusion limit.
Thus, in order to simplify the analysis, in what
follows, we will ignore the transition from weak to strong diffusion case and 
set
$\tesc(E)\propto \tsc(E)$. 

\begin{figure}[!ht]
\begin{center}
\includegraphics[width=8.0cm]{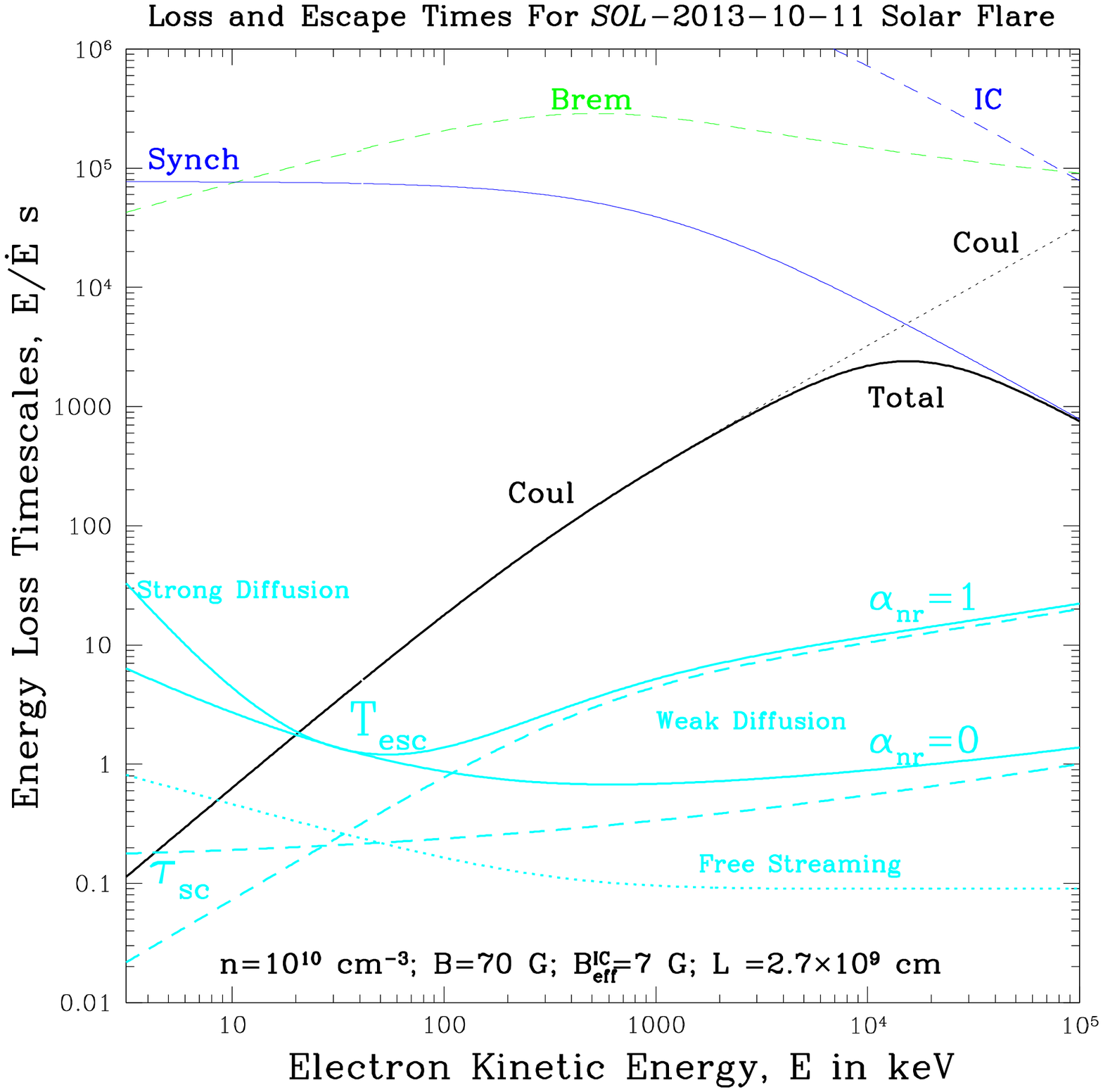}
\includegraphics[width=8.0cm]{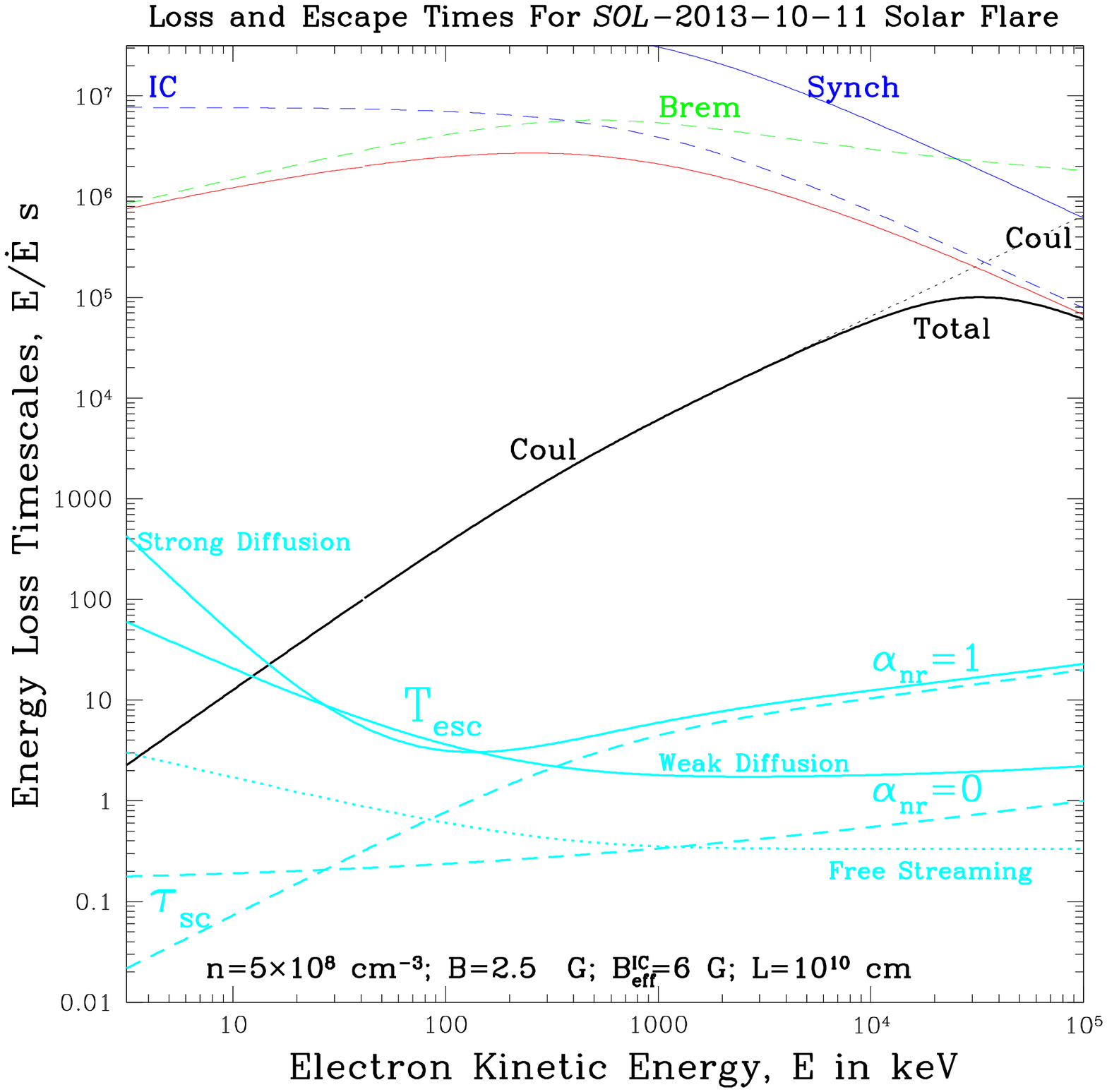}
\caption{Energy loss times for Coulomb (dashed-black),
Bremstrahlung (dashed-green), synchrotron (solid-blue),  IC
 scattering by optical photons (dotted-blue), total radiative loss (red) and
total loss (solid-black). We use magnetic field values appropriate for these
flares based on the analysis of the radio data in \S \ref{sync}. In cyan we show
the crossing
(dotted), scattering (dashed) and escape (solid) times based on Eqs.
(\ref{tsc}) and (\ref{tesc}) (for $\eta=3$ and for two sets of parameters:
$\a_{\rm nr}=1.0,
\tsc_0=3.0$ and  $\a_{\rm nr}=0.0, \tsc_0=0.3$) showing transition from strong
to weak diffusion at $\tsc=\tcross$.
Note that,  for energies of interest here (30 keV to 100 MeV), energy
losses
can be neglected and we have a thin target situation.}
\label{fig:times}
\end{center}
\end{figure}

In what follows we will deal  mainly with  $\nufnu$ spectra integrated over the
LT source region and the specified duration $\Delta T$ around the
peak of the impulsive phase
emission so that
$Q(E)=\int_{\Delta T} {\dot Q}(E,t) dt$  (or $N(E)=Q/\tesc(E)$) is the
total number
of injected (or accelerated) particles, and 
$\tesc(E)$  will be the time averaged escape time.

\subsection{Electron Bremsstrahlung and HXRs}
\label{brem}

The NTB $\nufnu$ spectrum of photons with energy $\e$ (in units of
$m_ec^2$)
produced by nonthermal electrons (interacting with background ions  at
nonrelativistic energies but with both electrons and ions in the relativistic
regime) is obtained using differential cross section (integrated
over angles) $d\sigma/d\e$ [see Eq.~(3BN) of Koch \& Motz (1959; KM59)] as:
\beq
J(\e) = {1\over \e\tbr}\int_{\e}^\infty {\tesc(E) f(\e, E)\over \beta(E)}
Q(E) dE,
\label{bremspec}
\eeq
where $\beta=v/c$, and 
\beq
\tbr=3/(16 \alpha r_0^2cn_{\rm eff})=1.1\times 10^{5}(10^{10}{\rm
cm}^{-3}/n_{\rm eff}) {\rm s}
\label{taubrem}
\eeq
for $\alpha=1/137$ and $r_0=e^2/m_ec^2=2.8\times 10^{-13}$ cm.%
\footnote{For fully ionized plasma the effective density  $n_{\rm
eff,nr}=\Sigma_i (Z_i^2n_i)$ at nonrelativistic energies and $n_{\rm
eff,rel}=n_{\rm eff,nr}+n_e$, where $n_e=(1+X)/(2X)n_p$ and $n_i=X_i/(XA_i)n_p$.
Here  $Z_i, A_i$ and $X_i$  are the charge,
atomic number and fractional mass
of ions, and $n_p$ is the proton density with $X_1=X$. The
background densities are usually obtained
from the emission measure of the thermal HXR component, $EM=V\Sigma_i
Z_i^2n_in_e=Vn_{\rm eff,nr}n_e$.  For
solar abundances ($X_1=X\sim 0.74, X_2=Y\sim 0.25$ and $Z=\Sigma_{i>2} X_i\sim
0.01$) it is easy to show that $n_{\rm eff,nr}=2\sqrt {EM/V}/(1+X)\sim 1.1\sqrt
{EM/V}$ and $n_e=\sqrt {EM/V}/[1+Z(-1+{\bar Z^2_i/A_i}]$. So that for an
average $Z_i^2/A_i\sim 4$ we get 
$n_{\rm eff,rel}=2.1\sqrt {EM/V}$. In what
follows we will use $n_{\rm eff}=(1$ and $2)\sqrt {EM/V}$ for
nonrelativistic
and
relativistic regimes, respectively.} 
The function $f(\epsilon, E)$ is a complicated but   slowly
varying function of  $x=\e/E$; in the
nonrelativistic regime
$f_{\rm nr}(x)=\ln {1+\sqrt{1+x}\over 1-\sqrt{1+x}}$ and for extreme
relativistic regime
$f_{\rm er}(x)=(1-x+3x^2/4)[\ln E+0.19 -\ln (x^{-1}-1)]$. (In our numerical
calculations we will use the
exact cross section in KM59).
For relativistic energies we add contribution of electrons as described in 
footnote 5 (see also Appendix A).

 Thus, given
the $\tesc(E)$ as described above we can use Eq. (\ref{bremspec})  to obtain the
total flux (in and out of the
 source), $Q(E)$, of
the accelerated electrons during the impulsive phase.
In general, for most solar flares the  HXR spectra $J(\e)$ decrease
rapidly with energy (power-law index $\g_X >3$), with
most of the emission in the nonrelativistic regime, so that the
lowest
energy 
value of  $\e_0^2J(\e_0)$ provides a good estimate of the {\bf total photon
energy} integrated over the duration $\D T$ of the flare:
\beq
\label{totalphoton}
{\cal E}_{\rm NTB}(>\e_0)=C\int_{\e_0}^\infty \e
J(\e)d\e=C{\e_0^2J(\e_0)\over \g_X-2}, 
\eeq
 where $C=4\pi d^2\D T$ and $4\pi d^2=2.8\times 10^{27}$ cm$^2$
for distance $d$ of 1 AU. This is the case
for Oct13 flare with $\g_X=3.2$ but not Sep14 where the HXR  $\nufnu$ spectrum
is flat
(i.e.~$\g_X\sim 2$ or $\e^2J(\e)\sim$ const.) over several decades in energy,
$\Delta \ln
\e\sim 7$ giving the total photon energy ${\cal E}_{\rm brem}(>\e_0)\propto 
\e_0^2J(\e_0)\Delta \ln \e$. In what follows we relate the photon energies to
the total flux
and  energy of electrons $Q(>E_0)=\int_{E_0}^\infty Q(E)dE$ and ${\cal
E}_e(>E_0)=\int_{E_0}^\infty EQ(E)dE$.

\vspace{0.3cm}

{\bf SOL2013-10-11:} As mentioned above the Oct13 flare has a 
 well defined nonthermal spectrum, a simple power law with $\g_X=3.2$, 
between 30 and 100 keV (based on \r and {\it Fermi}-GBM data). Thus,  we can use
the nonrelativistic
approximations
($\beta^2\sim 2E, \tesc=T_{\rm esc,0}E^{\alpha_{\rm nr}}$,  $f_{\rm
nr}$) and for a
power law electron spectrum $Q(E)=Q_0E^{-\d}$ obtain (see, e.g.~Lin \&
Hudson 1971; Brown 1972; Petrosian 1973) 

\beq
\label{bremOct13}
J(\e) = J_0\e^{-\g_X}\,\,\,\, {\rm with}\,\,\,\, J_0={T_{\rm esc,0}Q_0\over
\sqrt{2}\tbr}I(\g_X-1)
\eeq
and $\g_X=\d+0.5-\a_{\rm nr}$, where
\beq
I(n)=\int_0^1 x^nf_{\rm nr}(x)dx={\Gamma(n+1)\Gamma(1/2)\over
(n+1)\Gamma(n+3/2)},
\label{integral}
\eeq
where $\Gamma$'s stand for the gamma function. From these and  the observed
photon flux $J_0$ and $\g_X=3.2$,
we can obtain injected electron flux  at $E=mc^2$
\beq\label{qu0}
Q_0=J_0{\sqrt{2}\tbr\over [T_{\rm esc,0}I(2.2)}=10^5J_0\eta_O,
\eeq
where we have defined $\eta_O\equiv {10^{10}{\rm cm}^{-3}\over
n_{\rm eff}}{5{\rm s}\over T_{\rm esc,0}}$,
or the average accelerated electron  spectrum
\beq
\label{espec}
N(E)=Q(E)\tesc(E)={\sqrt{2}\tbr\over I(\g_X-1)}J_0E^{-\g_X+0.5}.
\eeq

\begin{figure}[!ht]
\begin{center}
\includegraphics[width=8.0cm]{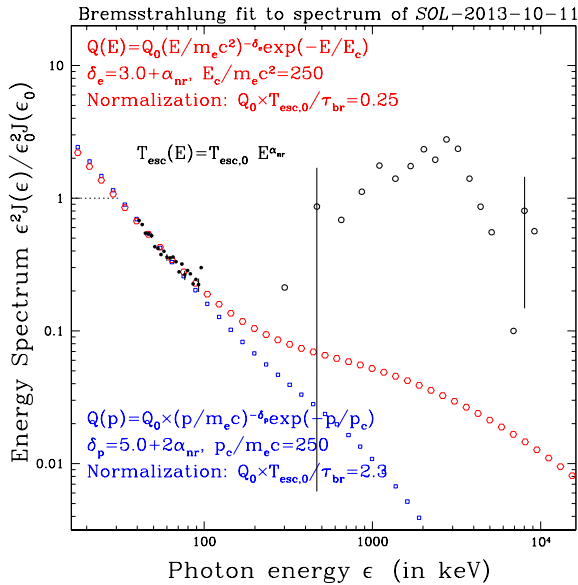}
\caption{Comparison of the calculated and observed HXR photon spectra
integrated for the
duration $\D T=25$ min (from 7:10 to 7:35 UT) Observations are given by filled
circles 
with some representative error bars; the open circle except two with error bars
are all upper limits (from Ack17). The calculated  spectra  are for power law
electron flux in momentum (blue) and energy (red) with a high
energy cut off, and for the shown escape time parameters. Note that
$\d_e=(\d_p+1)/2$ as
expected for nonrelativistic regime.  The calculated spectra are based on the
numerical integration of Eq.~(\ref{bremspec}) with the exact function $f(\e,E)$
from KM59. $\e_0^2J(\e_0)=9\times 10^{-8}$ erg cm$^{-2}$ s$^{-1}$
for $\e_0=0.059$, or 30 keV, as observed (see Table 1).
Note that $0<\a_{\rm nr}<1$ is a free parameter and the calculated fluxes in
the observed range  are not affected by the relativistic index $\a_{\rm
er}$ and
the cutoff energy (or momentum).}
\label{fig:OctHXRFits}
\end{center}
\end{figure}

Note that the energy dependence of the escape time does not enter in
the determination of the spectrum of the accelerated electrons, $\d_N\equiv -d
\ln N/d\ln E= \g_X-0.5=2.7$, while the index of the total flux $Q$ is 
$\d=2.7+\a_{\rm nr}$ so for the range
$0<\a_{\rm nr}<1.0$ we have $2.7<\d<3.7$.  
Fig.~\ref{fig:OctHXRFits}\ shows the calculated photon spectra for simple power
law
electron spectra
(in both momentum and energy) for the  flux $Q$ (with very high
energy exponential cutoff not relevant here) and for escape time index $\a_{\rm
nr}$ as a free parameter. 
(We use the exact bremsstrahlung cross section; formula 3BN, KM59).
As expected, in the nonrelativistic range the two spectra agree with each other
(but, of course,  with different indexes) and with the
observations based on {\it Fermi}-GBM data (which agrees with \r data for
this flare). However, the two model spectra
begin to diverge in the relativistic regime (with the power law in energy
predicting higher emission). These deviations are beyond the observed HXR range,
where we have only upper limits (open circles;
(except two possible detection with large error
bars).  As shown below radio observations shed light on the spectra at these
energies. Note also  that the values of index obtained from numerical fits,
$\d_e=\a_{\rm nr}+3.0\pm 0.05$ (or $\d_p=2\a_{\rm nr}+5.0\pm 0.05$) are slightly
different than the above values ($\d_e=2.7+ \a_{\rm nr}$) which assume the
nonrelativistic approximations (e.g.~$\d\ln \beta/d\ln E=1/[(E+1)(E+2)]\sim
0.5$)

Using these fit parameters, the observed $\e_0^2J(\e_0)=9\times 10^{-8}$
erg cm$^{-2}$ s$^{-1}$ at $\e_0=0.06$, we can obtain the total
number,
and energy  of the injected (or escaping)
electrons for the $\Delta T=25$ min duration of the flare as follows.
Using the normalization value (for energy fit) of 0.25 shown in
Fig.~\ref{fig:OctHXRFits} we get%
\footnote{If we use the approximate nonrelativistic relations in
Eqs.~(\ref{bremOct13}) and (\ref{qu0}) and the fact that
$J_0=C[\e_0^2J(\e_0)]\e_0^{\g_X-2}/(mc^2)$ we get the  normalization
$[\sqrt{2}/I(2.2)]\e_0^{1.2}=0.15$ instead of 0.25.} 

\beq\label{qu0Oct}
Q_0=0.25{\tbr\over T_{\rm esc,0}}{C\e_0^2J(\e_0)\over mc^2}=2.5\times
10^{34}\eta_0,
\eeq

Given $Q_0$ we then obtain  the total electron number and energy (above
$E_0=30$ keV) as:
\beq\label{numberOct}
Q(>E_0)=Q_0{E_0^{\a_{\rm nr}-2}\over \d-\a_{\rm nr}-1}=3.6\times
10^{36}\eta_0
\eeq
and
\beq\label{energyOct}
{\cal E}_e(>E_0)=Q_0{E_0^{\a_{\rm nr}-1}\over \d-\a_{\rm
nr}-2}=3.4\times 10^{29}\eta_0 \,\,{\rm erg}, 
\eeq
where we have used the fitted index $\d_e=3-\a_{\rm nr}$, with $\a_{\rm
nr}=0.0$. For  $\a_{\rm nr}=0.5$ the number and energy values will be larger  
by factors of 5.5 and 8.2, respectively.

\vspace{0.3cm}

\begin{figure}[!ht]
\begin{center}
\includegraphics[width=8.0cm]{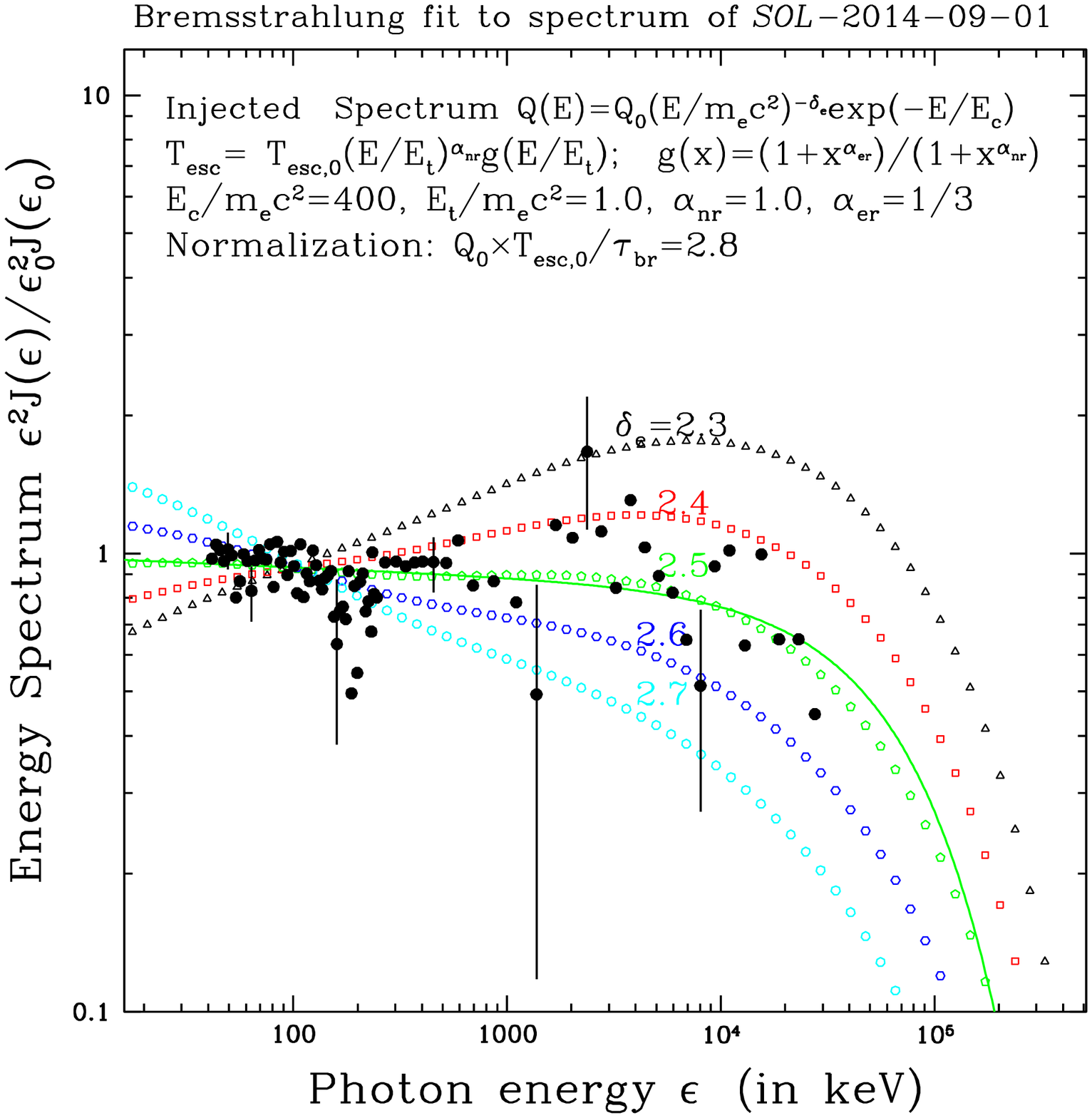}
\includegraphics[width=8.0cm]{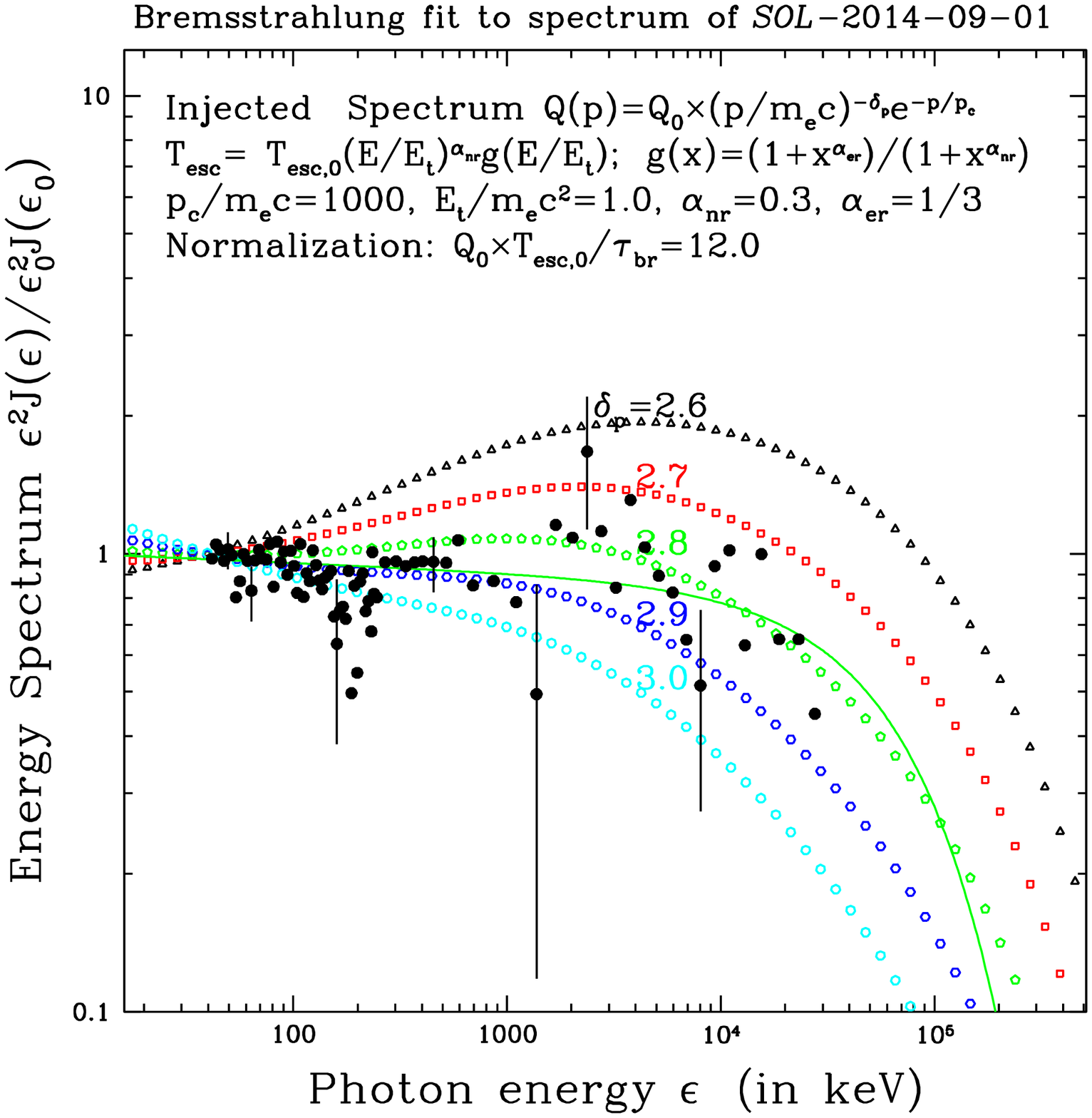}
\caption{Same as Fig.~\ref{fig:OctHXRFits} but fits to electron  energy (top)
and momentum (bottom) spectra separately, and  with $\e_0^2J(\e_0)=4\times
10^{-7}$ erg cm$^{-2}$ s$^{-1}$ for $\e_0=0.059$, or 30 keV. The green curves
 show $\e^2J(\e)\propto \e^{0.02}\exp {\e/180}$, the fit function in
Ack17. Observations represent the spectra integrated for the
duration $\D T=18$ min from 11:02 to 11:20 UT (from Ack17). }
\label{fig:SepHXRFits}
\end{center}
\end{figure}

{\bf {\it SOL}2014-09-01:} We can carry out a similar analysis  for Sep14
flare as well. However, because here we have a nearly flat $\nufnu$ flux
extending over three decades in energy from nonrelativistic
to extreme relativistic
regime (30 keV to 30 MeV) we need to rely on numerical solutions. In fact as
shown in appendix A
it is difficult to obtain such a spectrum via bremsstrahlung emission because
of the changes in the energy-momentum-velocity relation and the
bremsstrahlung cross section across the
trans-relativistic region. For  a simple power law  spectrum of the  accelerated
particles, 
$N(E)=Q(E)\tesc(E)\propto E^{-\d_{Ne}}$, and using the nonrelativistic and
extreme
relativistic forms of the function $f(\e, E)$ in Eq.~(\ref{bremspec}),
it is easy to show that one obtains, respectively,  photon spectra $J_{\rm
nr}(\e)\propto \e^{-(\d_{Ne}+0.5)}$ and $J_{\rm er}\propto \e^{-\d_{Ne}}(\ln
\e+c_1)$
(with $c_1$  a constant of order unity), which indicates spectral hardening
of $\sqrt{\e}\ln \e$ or photon index change of
$\g^{\rm nr}_X-\g^{\rm er}_X=0.5+1/(\ln
\e+c_1)$. Thus, to get a power law photon spectrum we need a BPL
spectrum of accelerated electrons, $N(E)$, that
steepens for $E>1$. However, this spectral hardening  can be compensated by a
break in $\tesc(E)$, which, as can be seen from Eq.~(\ref{tsc}), is the
case for $\a_{\rm nr} > \a_{\rm er}=1/3$, so that a simple power law  of
injected
electrons
$Q(E)$ can reproduce  the observations. As shown in the top panel
of Fig.~\ref{fig:SepHXRFits}, this is the case for $\a_{\rm nr}=1.0, \a_{\rm
er}=1/3$ and $\d_e=2.5\pm 0.1$.

Similarly, for a simple power law  in
momentum, $N(p)\propto p^{-\d_{Np}}$, we
have $J_{\rm nr}(\e)\propto \e^{-(1+\d_{Np}/2)}$ and $J_{\rm er}\propto
\e^{-\d_{Np}}(\ln \e+c_1)$; again with spectral index $\g_X$ changing from
$1+\d_p/2$ to $\d_p-1/(\ln \e+c_1)$. In this case we have a spectral softening
(or steepening) for $\d_p>3$ (which is usually the case). Thus, in momentum
space we need an electron spectrum that gets harder (flattens) in the
relativistic range. However, as shown in Appendix A,  for $2.3<\d_p<2.5$ the
logarithmic part can compensate for
this steepening and give a nearly flat  $\e^2J(\e)$
spectrum across the
trans-relativistic range. 
Again, for the injected (or escaping) spectrum, $Q(p)$, we need to include the
energy dependence of $\tesc$. The above discussion implies that we need a weaker
(or no) energy dependence for $\tesc$. As shown in the bottom panel of
Fig.~\ref{fig:SepHXRFits}, we obtain acceptable fits for $\a_{\rm nr}=0.3$ and
$\d_p=2.8\pm 0.1$. 

In summary, power-law  injected
spectra (with exponential cut off at above few 100 MeV) both in momentum and
energy space can explain the observations with different values of index
$\a_{\rm nr}$ but
well within the range obtained empirically by CP13 and P16. Note however
that, if we include the transition from weak to strong diffusion the photon
spectra
will be steeper than shown in the above figures at (low) energies below the
observed range. 

Following the same procedure as above, we can also derive the  total number and
energy flux of the electrons. We will use the fit parameters in the energy
space which is simpler. The fitted index $\d_e=2.5$ and normalization $Q_0T_{\rm
esc,0}/\tbr=2.8$ used in obtaining the
fit (top panel Fig.~\ref{fig:SepHXRFits}) implies 
\beq\label{qu0Sep}
Q_0=2.8{\tbr\over T_{\rm esc,0}}{C\e_0^2J(\e_0)\over mc^2}=4.5\times
 10^{35}\eta_S
\eeq
where we  set $\e_0^2J(\e_0)=4\times 10^{-7}$ erg cm$^{-2}$ s$^{-1}$), $C=4\pi
d^2\D
T=3.0\times 10^{30}$ cm$^{2}$ s (for duration $\D T=18$ min), and we have
defined $\eta_S\equiv {10^9{\rm cm}^{-3}\over n_{\rm eff}}{10{\rm s}\over T_{\rm
esc,0}}$. From this we can get the total number and energy flux of injected
electrons above energy $E_0=0.059$ (30 keV) as
\beq
\label{NumEnergySep}
Q(>E_0)= 2.1\times 10^{37}\eta_S\,\,\,{\rm and}\,\,\,
{\cal E}_e(>E_0)=3.0\times 10^{30}\eta_S \, {\rm erg}.
\eeq
Here we have ignored the exponential
cut off which will reduce
these numbers by a factor $<1-\sqrt{E_c/E_0}\sim 0.99$.

\subsection{Electron Synchrotron and Microwaves}
\label{sync}

\subsubsection{General Synchrotron Spectra}

\begin{figure}[!ht]
\begin{center}
\includegraphics[width=8.0cm]{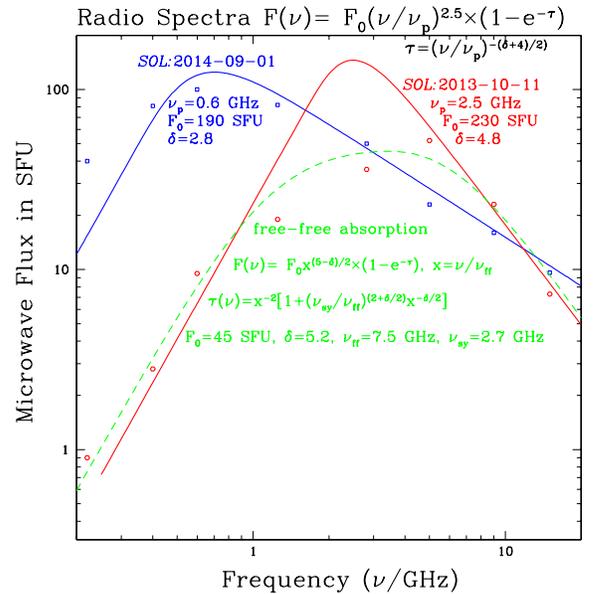}
\caption{Observed (points; from Ack17) and self-absorbed fitted (curves)
spectra for Oct13
(red) and Sep14 (blue) BTL flares. The dashed green curve includes free-free
absorption, which provides a better fit for Oct13 flare (see Appendix B).}
\label{fig:radio}
\end{center}
\end{figure}

Fig.~\ref{fig:radio} shows the observed microwave spectra (points) of the two
flares. Synchrotron emission by relativistic electrons is the most
likely mechanism of these emissions. The high frequency optically thin
portion is observed
over only one decade ($\sim 1<\nu<\sim 10$ GHz, with the spectral index $\g_r$)
so that only a fit to a simple power law  electron density
spectrum ($n(E)=n_0E^{-\d}$) is possible.
In addition, because of the 
unusually large height (above the photosphere) of these sources, we most likely
are dealing with  lower than usual magnetic fields, 
lower gyro-frequencies, $\nu_B=2.8\times 10^6 (B$/G),  high harmonic
$\nu/\nu_B>(300{\rm G})/B$ and Lorentz factor ($\g \sim 14
\sqrt{{\rm G}/B}$).
Thus, we are most likely in the relativistic
regime, with no difference between the spectra in energy and momentum
spaces, and we can
use the usual relativistic formulation of the synchrotron emission and
absorption coefficients $J(\nu)$ and $\kappa(\nu)$ (see, e.g.~Rybicki \&
Lightman 1979):
\beq
\label{remissivity}
J(\nu)= \a h\nu_Bn_0a(\d)x^{(1-\d)/2}, \,\,\, {\rm with}\,\,\, x=\nu/\nu_B,
\eeq
and 
\beq\label{absorbtion}
\kappa(\nu)=\a {n_0\over 4\pi} \left({h\nu_B\over mc^2}\right)
\left({c\over \nu}\right)^2b(\d)x^{-\d_/2},
\eeq
where $a(\d)$ and $b(\d)$ are slowly varying functions of order unity (see
Appendix B).
From these we get the source term
\beq\label{source}
S(\nu)={J(\nu)\over 4\pi \kappa(\nu)}=
m\nu_B^2c(\d)x^{5/2},\,\,\, {\rm with}\,\,\, c(\d)=a(\d)/b(\d),
\eeq
and the spatially integrated  radio flux
$F(\nu)= S(\nu) \Omega f(\tau_\nu)$,
where $ \Omega$ is the angular size (in sr) and 
\beq\label{opticaldepth}
\tau_\nu=\int \kappa dl\simeq \tau_0x^{-2-\d/2}\,\,\, {\rm with}\,\,\,
\tau_0\equiv {\a n_0Lb(\d)\over 4\pi}{h\over
m\nu_B},
\eeq
is the optical depth (integrated over the source depth along the line of
sight, $L=V/A$).
The function $f(\tau)$ depends on
the source shape and size, and magnetic field geometry. However, as shown in
Appendix B, the spatially integrated results depend weakly on the exact form of
this function. 
In what follows we will use primarily the plane parallel radiative transfer
relation $f(\tau)=1-e^{-\tau}$. 
In general however, in the optically thin ($\tau_\nu\ll 1$) regime
$f(\tau)=\tau$ and 
\beq\label{thinF}
F(\nu)=\a n_0L(\D \Omega/4\pi)h\nu_Ba(\d)x^{(1-\d)/2},
\eeq
and in the optically thick ($\tau_\nu\gg 1$) regime
$f(\tau)=1$ and we get 
\beq\label{thickF}
F(\nu)=(\D \Omega/4\pi)m \nu_B^2c(\d)x^{5/2}, 
\eeq
with a peak flux $F_p=S(\nu_p) \Omega f(\tau_p\sim 1)$, at frequency
$\nu_p$. 

\subsubsection{Electron Characteristics}

From the observed spectral index $\g_r$ of microwave flux in the  optically
thin regime, we determine the electron index $\d=2\g_r+1$ (and hence $a(\d),
b(\d)$).%
\footnote{This and all of the above relativistic relations are  valid for low
magnetic fields ($\nu_B\ll \nu$) (and hence high Lorentz factors $\g\sim
\sqrt{\nu/\nu_B}$). As shown in Petrosian (1981) (see also
Petrosian \& McTiernan, 1983), in the semi-relativistic regime these relations
are
more complicated. In general, for a power law electron index the synchrotron
spectra steepen at lower frequencies (see, e.g.~Ramaty, 1969), so that the
relation between $\d$ and $\g_r$ varies slowly with frequency (see,  Ramaty \&
Petrosian, 1972).  Using numerical results, Dulk  (1985) gives the
semi-relativistic relation $\d\sim 1.11\g_r+1.36$, which is an approximate
average value.}

\begin{deluxetable*}{cccccc}[ht]
\label{tab:indexes}
  \tablewidth{\textwidth}
  \tablecaption{Electron Indexes and Numbers for flux $Q$}
  \tablehead{
  \colhead{Flare} & \colhead{$\d_e$  HXR} &  \colhead{$\d_p$ HXR} &
\colhead{$\d_e=\d_p$ Radio} & \colhead{$Q_0$ HXR} & \colhead{$Q_0$ Radio} 
}\\
\startdata
${\it SOL}2013-10-11$ & $3.0+\a_{\rm nr}$ & $5.0+2\a_{\rm nr}$  &
$(5.1-5.5)$ & $2.5\times 10^{34}$ & $5.5\times 10^{32}$\\
${\it SOL}2014-09-01$& $2.5;\, (\a_{\rm nr}=1.0)$ & $2.8; \, (\a_{\rm nr}=0.3)$ 
& $3.1$ 
& $4.5\times 10^{35}$ & $4.4\times 10^{35}$\\
\enddata
 \end{deluxetable*}
 
\vspace{0.3cm}

As is well known
flux measurements in this regime is not
sufficient to determine the number (or energy) of the electrons because of the
degeneracy between  $n_0$ and magnetic field $B$ (or
$\nu_B$). Observations in the optically thick  regime [Eq.~(\ref{thickF})]
provide the second datum
which allows us to break this degeneracy and determine both $n_0$ and $B$.
Using the expression for the source in Eq.~(\ref{source}), it is easy to show
that we can write (see Appendix B for  more details)
\beq\label{nub}
\nu_B\simeq [c(\d) f(\tau_p)]^2\left({m \Omega
\over F_p}\right)^2\nu_p^5,
\eeq
which then can be used in Eq.~(\ref{thinF}) along with flux measurements in the
optically thin regime to determine $n_0$, or the spatially
and temporally integrated number $N_0=\int dt \int n_0({\vec r},t) dV$ as
\beq\label{numberR}
N_0=n_0V\Delta T={C{\bar F}(\nu)\over \a a(\d)h\nu_B}x^{(\d-1)/2},
\eeq
where we have used $V/(L\D \Omega)=d^2$  and ${\bar F}$ is the average flux for
the duration of the microwave flare. The spectra shown in
Fig.~\ref{fig:radio} are for about one minute duration around the peak of the
radio light curve. For the purpose of comparing with electron numbers and
spectra obtained from the analysis of the NTB emission, we need the value of
flux averaged over the same durations used above ($\D T=25$ and 18 min for
Oct13 and Sep14 flares, respectively). Since the radio light curve are almost
triangular (see Figs.~2 and 5 in Ack17), we estimate average fluxes 
of 1/2 and 3/4  of the peak-time fluxes shown
in Fig.~\ref{fig:radio} and given in Table 1, for Oct13 and Sep14,
respectively.

In Fig.~\ref{fig:radio} we show self-absorbed spectra based on the above
equations
superimposed on the RSTN observations of the two flares from which we can
determine $\nu_p$ and $F_p$. These are not very accurate fits, especially for
Oct13 flare, but allow us to
obtain a rough estimates of the required quantities. In particular,
the value of $B$ thus obtain is very uncertain for several reasons. One, as
evident from Eq.~\ref{nub}, $\nu_B$ is very sensitive to the
measured parameters; it depends on the fifth
power of  $\nu_p$ and square of $F_p$.
Two, inhomogeneities in the
source can bias the result. Three, there may be other absorption
processes, in particular as shown by Ramaty \& Petrosian (1972) free-free 
absorption may be important in a high elevation, low magnetic field situation.
In
fact, the spectrum of the Oct13 flare in Fig.~\ref{fig:radio} shows some
flattening around 5 GHz, perhaps due to free-free absorption, with possible
emergence of self-absorption around 1 GHz. As
described in Appendix B, and shown by the dashed green curve, inclusion of
free-free absorption improves the fit considerably. As also indicated in
Appendix B, this model also implies presence of optically thin  free-free
emission from 5 GHz to soft X-rays  of $<10$ keV  well below the observed
microwave fluxes and in rough agreement with the thermal
bremsstrahlung flux observed below 10 keV (see, Pesce-Rollins et al. 2015).

In Appendix B
using a self-absorbed model  for the Sep14  flare  we obtain magnetic field
values ranging from $2-20$ G. The fact that
for this flare with a height of $10^{10}$ cm  we get magnetic fields lower than
the usual $B \sim 100$ G associated with low lying ($\sim 10^9$ cm) LT sources
is
encouraging. A self-absorbed fit to Oct13 flare gives $B$ values in the range
300 to 3000 G. This is most likely not correct because of the poor fit. Using
the fit parameters including free-free absorption  yields a more reasonable
value of $\sim 200$ G. We use these values of $B$ (or $\nu_B$) and the fluxes at
$\nu=10$ GHz (in the optically thin range) in Eq.~(\ref{numberR}) to
calculate  the number of electrons required for the production of the
microwaves. 
 For the Oct13 flare with fit parameters $\d=5.2$, $a(\d)=2.6$  and $B=200$ G
(obtained from the fit including free-free absorption)  and the observed flux 
${\bar F}(\nu=10$ GHz)=10 SFU we obtain $N_0$ or $Q_0=N_0/T_{\rm esc,0}$ to be
\beq\label{qu0ROct}
Q_0=5.5\times 10^{32}\left(5{\rm s}\over T_{\rm esc,0}\right)\left(200 {\rm
G}\over B\right)^{3.1}.
\eeq
This should be compared with $2.5\times 10^{34}$ obtained in Eq.~(\ref{qu0Oct}).
There are however two uncertain parameters; $n_{\rm eff}$ and $B$. For
example the two estimates would agree for $n_{\rm eff}=5\times 10^{10}$
cm$^{-3}$ and $B=70$ G. Note that for this magnetic field $\nu_B=0.2$ GHz, 
and the lowest
reliable observed microwave point  of $\nu=0.6$ GHz is  produced roughly by
electrons  with
Lorentz factor $\g\sim (\nu/\nu_B)^{1/2}=1.7$ so that relativistic
expressions used here begin to break
down and one should use the semi-relativistic expressions. However, at such low
frequencies we are in the optically thick regime, while the values of $B$ and
$n_0$
are determined by higher frequency data points.
For the Sep14 flare using self-absorbed fit parameters $\d=2.7$, $B=2.5$ G,
$a(\d)=0.1$  and ${\bar F}(\nu=10$ GHz)=10 SFU we obtain 
\beq\label{qu0RSep}
Q_0=4.4\times 10^{35}\left(10{\rm s}\over T_{\rm esc,0}\right)\left(2.5 {\rm
G}\over B\right)^{1.85},
\eeq
which is somewhat  fortuitously exactly what was obtained from X- and gamma-ray
observations given  in Eq.~(\ref{qu0Sep}).

\subsection{Combined Electron Spectra}
\label{electrons}

\begin{figure}[!ht]
\begin{center}
\includegraphics[width=8.0cm]{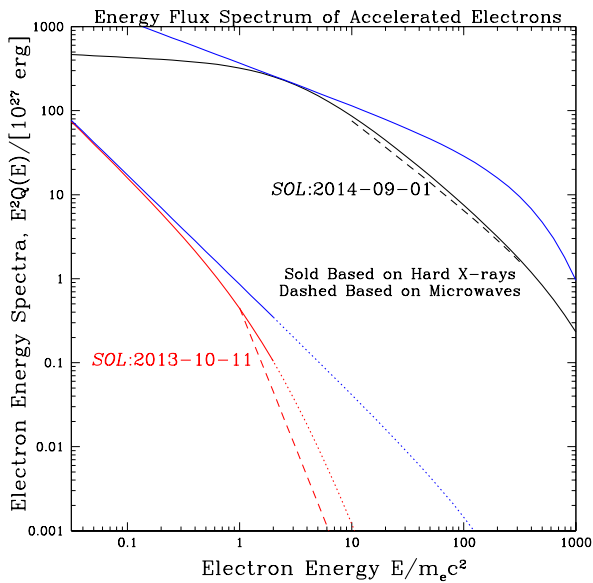}
\caption{Total energy flux spectra of accelerated electrons as power laws in
momentum space with exponential cut off, for
Oct13
(red) assuming $B=70$ G and Sep14 (black) assuming $B=2.5$ G, using spectral
parameters obtained from fits to HXR
data (solid) and radio data (dashed). As evident there is excellent
agreement between radio and X-ray based spectra. Blue curves use parameters
based on
fits to HXRs and  power law in energy space, which shows deviation from
spectra based on radio data. Dotted sections are extrapolations.}
\label{fig:electrons}
\end{center}
\end{figure}

We now combine the results obtained for the electron characteristics from HXR
and microwave data. In Table 3 we summarize our results on electron spectral
indexes assuming  Kolmogorov
turbulence with $\a_{\rm er}=1/3$. For the Oct13 flare the index of
$5.1-5.5$ obtained from the microwave data agrees with momentum index based
on HXRs with $\a_{\rm nr}=0.05-0.25$, but agreement with the energy index
requires
either an unusually large $\a_{\rm nr}>2$ or a spectral steepening (by 1 to 2
units) above  $E\sim mc^2$.
For the Sep14 flare the radio index of 2.8 for $N(E)$ and $\sim 3.1\pm 0.1$ for $Q(E)$ is closer to
the HXR
momentum
index
of $2.8\pm 0.1$ than the energy index of $2.5\pm 0.1$. For this flare the
values of $Q_0$ (or number of electrons at $E=mc^2$) obtained from radio
and x-gamma-ray data are in excellent agreement. But as mentioned above
some adjustments of uncertain parameter values (such as $n_{\rm eff}$, $\tesc$,
etc.) is needed for an acceptable agreement for the Oct13 flare. Figure
\ref{fig:electrons}
summarizes these findings. 

In addition to the preliminary analysis in Ack17
mentioned at the outset, there have been similar determination of electron
spectra based on HXRs (Share et al 2017; Plotnikov et al. 2017)
assuming  both thin and thick-target, based only on electron energy spectra, and without
consideration of the energy dependence of the escape time. As expected
the
electron indexes derived in these papers  are different than those presented
here, which not only are for  a thin target model but also include the energy
dependence of the escape time. And in the case of Sep14 flare the analysis here
includes the exact relativistic bremsstrahlung cross section. These factors can
account for such differences.

\subsection{Emissions by Escaping Electrons }

Some of the particles escape  along
closed field lines to the FPs to the  AR located BTL and visible onle to \stereo.  They
lose all their energy at the FPs and
produce {\bf thick target} HXRs and microwaves. Some escape out of the corona along open
field lines and eventually reach the Earth and are detected as SEPs by
near-Earth instruments. As shown in P16, the escape times up, $T_{\rm
esc}^u(E)$, and
down, $T_{\rm esc}^d(E)$, will most likely have different values and
energy dependences,  so that the flux of SEPs will be different than those
traveling to the FPs and produce HXRs. As shown in Krucker et al. (2007), 
observations indicate that
most of the particles are directed downward and produce thick-target HXR and
microwave emission more
efficiently
than in the LT region. For example the NTB spectrum would be  

\beq
J_{FP}(\e) = {1\over \e\tbr}\int_{\e}^\infty 
{\tloss(E) f(\e, E)\over \b(E)} dE\left({1\over E}\int_E^\infty Q(E')dE'\right),
\label{bremspecFP}
\eeq
which is similar to the thin target expression given in
Eq.~(\ref{bremspec}) but with two differences. The first
is that, instead
of $Q(E)$ we now have the effective electron spectrum given by the
integral in the parenthesis, which for a power-law injected spectrum is equal to
$Q(E)/(\d-1)$. The second is
that, instead of escape time, the integrand contains the energy loss time
$\tloss(E)=E/{\dot E}_L$ shown by the solid black
lines in Fig.~\ref{fig:times}. In the nonrelativistic limit (e.g.~for Oct13
flare) with $\tesc(E)\propto E^{\alpha_{\rm nr}}$ and $\tloss\propto E^{1.5}$
this will lead to a FP photon spectrum with index $\g^{\rm FP}_X=\d-1$ instead
of $\g^{\rm LT}_X=\d+1/2-\a_{\rm nr}$, implying that $\g^{\rm FP}_X=\g^{\rm
LT}_X-1.5+\a_{\rm nr}=1.7 +\a_{\rm nr}$. As shown in Fig.~\ref{fig:times} for
the energy range of 10 to few 100 keV $\tesc\sim $constant ($\a_{\rm nr}=0$) so
that the FP HXR emission will be much harder. Also, since loss time is about 10
times larger than the escape time in this energy range, the FP flux will be
correspondingly larger (modulo the factor $\d-1\sim 2$). These relations are
more complicated for Sep14 flare 
with HXRs extending into relativistic range, but in general we would expect
even a  harder and higher flux of FT emission. The same is true for
synchrotron emission by relativistic electrons where  one must also 
consider the synchrotron emission, absorption and loss process in  higher
magnetic fields at the FPs, which affect both the emission and energy loss
rates. This
implies that 10 to 100 times higher fluxes of HXRs
and microwaves are emitted from the FPs (in the AR BTL) than those emitted
from the LT.

The above equation is also applicable if the LT source was a thick  rather
than a thin-target source. As stated in \S 3 this will require an unusually
short scattering mean free path (i.e. short $\tsc$) or highly converging
magnetic field structure. but if these were the case it would require a steeper
accelerated electron spectra. For example, in the nonrelativistic HXR
emission case, instead of $\d_{\rm thin}=\g_X+\a_{\rm nr}-0.5$ [see discussion
related to Eqs.~(\ref{bremOct13}) and (\ref{integral})] one needs $\d_{\rm
thick}=\g_X+1$ which is steeper by (index higher by 1.5, for $\a_{\rm nr}\sim
0$). Similarly the required energy fluxes of electrons will be lower by a factor
equal to the average value of $\tesc(E)/\tloss(E)$  in the relevant energy
range. Thus, all the curves in Fig.~\ref{fig:electrons} would be lower and
steeper and the transitions from nonrelativistic to relativistic range would
be somewhat different.

\section{LAT GAMMA-RAYS AND ACCELERATED PROTONS}
\label{pion}

The {\it Fermi}-LAT emission of $>100$ MeV photons is different from the
impulsive emissions considered in this paper in two important ways. The
first difference is that centroids of the LAT sources are located
$\sim 65\arcsec$ and $275\arcsec$ away
from the centroids of the \r LT
sources for the Oct13 and Sep14 flares, respectively. 
The second is that, like  most flares detected by the {\it Fermi}-LAT, 
the LAT light curves of the flares under consideration are very different
than the light curves of  impulsive emissions. They rise somewhat later and
decay much
more slowly with a duration more similar to gradual SEPs that are believed to be
accelerated in the CME environment. Since {\it Fermi}-LAT flares are almost
always associated with fast CMEs, the possibility that the LAT emission is
produced by particles accelerated in the CME-shocks and escape from the shock
downstream toward the Sun has gained some momentum. For the BTL flares under
consideration here this scenario will require a magnetic connection between the
downstream region 
and areas in the photosphere in the visible disk far away from the AR where
these flares originated. Recent simulation (Jin et al. 2018; Plotnikov et al. 2017) indicate that
this is a likely scenario. 

These two differences point to a different origin  for the LAT
observations than the LT source considered above. However, based on the
localization data alone, the possibility that the LAT gamma-rays may also
be a thin-target emission coming from the LT \r location cannot be ruled out
with high
confidence. So it is important to explore this possibility as well. 
Just as in the case
of HXRs described above, a thin-target LT emission  would
require higher
energy contents for the accelerated protons by a factor equal to $\tloss/T_{\rm
esc}$ at the LT. The Coulomb loss time for 500 MeV protons is
$\tloss\sim 10^5$ s (for $n=10^{10} {\rm cm}^{-3}$), but, unlike for
electrons, we have no
empirically based information on the escape time. Assuming the same
(theory based) relativistic
approximation used for electrons, we estimate  escape times of 10 to 100 s.
This means that the production of the LAT gamma-rays at the LT would require
$10^4$ to $10^5$ times more energy for protons than that required for the
thick-target photospheric emission. 
Ack17, assuming  thick
target photospheric emission  
estimate proton energies of ${\cal E}_p(E>500$ MeV) $\sim 1$ and
$7\times
10^{25}$ erg, for Oct13 and Sep14 flares, respectively. This means that the LT
thin-target model would require proton energies in the range of  $10^{28-29}$
erg. These, though larger are still about 10 times
smaller than the energies
of
the
electrons shown in Fig.~\ref{fig:electrons}. The proton energies would become
comparable and could  exceed the electron energies if their spectra are
extrapolated to
10s of MeV. However, absence of a strong signature of nuclear de-excitation
lines
rules out this possibility. We therefore conclude that the possibility of a
thin
target LT source for gamma-rays cannot be ruled out with high confidence on
energetic grounds alone. {\it However, the differences in the light curves
and
centroids of HXR and $>100 $ MeV emissions favors a different acceleration site
and mechanisms for protons than HXR-microwave producing electrons.}

Finally, we consider the possibility of the LAT emission being due to electron
bremsstrahlung from a second relativistic  electron component with $0.1<E<5$
GeV. This component cannot be due to electrons accelerated at (and emitting
from) the LT source because they will produce microwaves of $\nu> 10$ GHz, and 
with
a flux much higher than that observed. On the other hand, if the emission comes
from the photosphere (produced, for example, by electrons that are
accelerated at the CME and
find their way to the photosphere) then such energetic electrons penetrate
to very
high densities just below the photosphere and lose almost all their energies
via bremsstrahlung emission. In that case the required  energy of electrons
would be slightly larger than the observed energies of $\g$-rays of $\sim
1.4\times 10^{24}$ and
$\sim 1.2\times 10^{25}$ ergs, for Oct13 and Sep14 flares, respectively) which
are
about
5 times lower than the  energy of proton  given in Ack17. {\it However,
acceleration of electrons to tens of GeV and their
transport over large distances requires a very high  acceleration or
a very low energy loss rate. This fact also  favors pion decay production of
$>100$ MeV photons.}

\section{SUMMARY AND CONCLUSIONS}

In this paper we present a detailed analysis of HXR and microwave spectra of
two solar flares (Oct13 and Sep14) which, based on \stereo observations,
originated 10 and 40 degrees BTL of the Sun, but were  detected by {\it Fermi},
{\it RHESSI}, {\it SDO} \konus and ground based radio telescopes. The relevant
observed characteristics are summarized  in Tables 1 and 2. The 20-30 min HXR
light curves observed by {\it RHESSI}, {\it Fermi}-GBM and \konus are almost
identical
and
similar to the radio light curves for both flares. The {\it Fermi}-LAT light
curves are somewhat delayed and last longer. \r images (up to 25-50 keV for
Oct13 and 6-12 keV for Sep14) show sources (of
size $\sim 50 \arcsec$) at the limb presumably the top of a relatively
large flare loop peeking over the limb. The LAT localizations puts the centroid
of gamma-ray emission  $65 \arcsec$ and $275 \arcsec$ away from the
\r
source
for Oct13 and Sep14 flares, respectively.

Based on the similarity of light curves we assume a co-spatial emission of HXR
and
microwave emissions and determine accelerated electron characteristics over a
broad range of energies from sub-relativistic regime (based on bremsstrahlung
emission of low HXRs) to extreme relativistic regime (based on bremsstrahlung
emission of gamma-rays and synchrotron emission of microwaves). In case of
Sep14 flare the measured electron bremsstrahlung emission extends from 30 keV to
$\sim
100$
MeV. This requires careful consideration of two important aspects. The first is
the question of the time accelerated particles spend in the source region, which
we call the escape time, and the second is that, because the observations  span
the
trans-relativistic region, we should distinguish between spectra in momentum and
energy space. Using empirically determined values and energy dependence of the
escape time
in 10-100 keV range  by CP13 and P16, and their extension to relativistic
energies based on theoretical considerations, we show that we are
dealing with a thin target processes which then allows us to get the electron
characteristics. Our results can be summarized as follows.

\begin{enumerate}

\item

From modeling of the NTB emission of Oct13 flare we find that simple power law 
electron
spectra in both momentum and energy space can reproduce the observed HXRs. For
Sep14 flare a simple power law  in momentum can describe the broad range
of the observed HXRs more
readily and with more reasonable values for the escape time index than simple
power law 
in energy. From these fits we determine the
spectral index,
numbers and energy content of accelerated electrons.

\item 

The radio spectra for both flares show a distinct optically thin emissions that
peak around 1 GHz and a well defined turnover at lower frequencies  indicating
emergence of a optically thick spectrum.  Self-absorbed
synchrotron spectrum provides an adequate fit for the Sep14 flare, but for
the Oct13 flare
a  self-absorbed synchrotron spectrum does not fit the observations in the
range $0.5<\nu<5$ GHz. We show
that a model whereby free-free absorption starts at about 7 GHz with
self-absorption becoming dominant below 2 GHz provides an acceptable fit. 
These modelings  allow us to determine both the spectrum and numbers of
relativistic electrons and
the magnetic field (that turn out to be lower than usual appropriate for the
large
height of the
source).

\item

We then compare the two electron spectra obtained by these two methods. We show
that for both flares  extrapolation of spectra based on HXRs to the relativistic
regime agree  with those based radio data assuming a simple power law  (with
exponential cut off at several 100 MeV) in momentum  but not in energy space.
The latter require a broken power law with a break at $E<mc^2$. The numbers and
energy content of these flares are in the right ball park and allow us to
predict the FP emissions from AR located BTL.  

\item 

We also consider the possibility of thin target LT emission of the LAT
gamma-rays and find that this requires 100 to 1000 time more energy of
accelerated protons compared to thick target photospheric emission. However,
even these energies are less than  those of the electrons so that this scenario
of high energy gamma-ray LT emission cannot be ruled out on energetic grounds.
This is also true for production of these higher energy gamma-rays by GeV
electrons at the photosphere. Nevertheless, because of the difficulty of
acceleration of electrons to several GeV, pion decay scenario is favored, and
the differences in the light curves and
centroids of HXR and $>100 $ MeV emissions indicates a different acceleration
site
and mechanisms for (pion producing) protons than (HXR-microwave producing)
electrons.

\item

The radiative signatures of occulted flares, such as those considered here,
provide the most direct information on spectra and energy content of
accelerated particles, and hence on the acceleration mechanism, uncontaminated
by the stronger FP emission. For example, the differences between the
required spectra in energy and momentum spaces can shed light on the details
of the acceleration process. This important aspect of the problem will be dealt
with in subsequent papers.

\end{enumerate}

{\bf Acknowledgements}: This work is
supported by NASA  LWS grant NNX13AF79G, H-SR grant NNX14AG03G and Fermi-GI
grant NNX12AO78G. I would  like to thank the {\it Fermi} colaboration, in
particular the corresponding authors of Ack17 (A. Allafort, M.
Pesce-Rollins, N. Omodei, F. Rubio and W. Liu) for help in preparation of this
paper. I
would also like to thank  anonymous referees for many helpful  comments.

\vspace{1cm}

\section{Appendix A: Some Aspects of Bremsstrahlung Emission}
\label{AA}

\begin{figure}[!ht]
\begin{center}
\includegraphics[width=8.0cm]{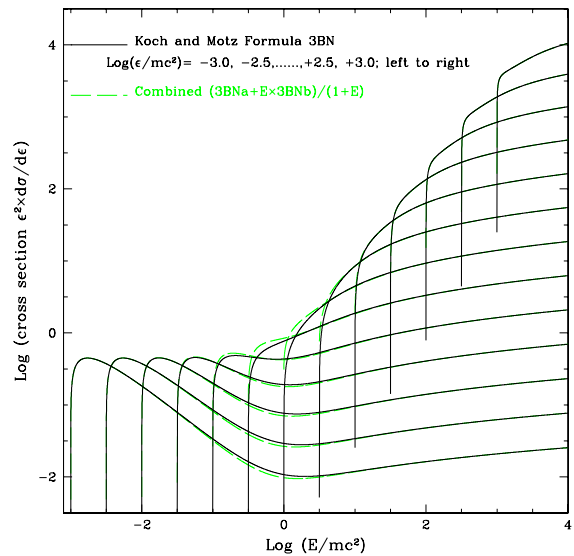}
\caption{Comparison of exact angularly averaged  bremsstrahlung cross
section (solid-black) using equation 3BN of KM59 with the approximate one in
Eq.~(\ref{bremcross}) (dashed-green) showing excellent agreement except for
deviations of less than a few percent around energies $\sim mc^2$.}
\label{A1}
\end{center}
\end{figure}

1. {\it Approximate Cross Section:}  The nonrelativistic and extreme
relativistic approximations given after Eq. (3) (same as expressions 3BNa and
3BNb of KM59, respectively) can be combined as
\beq\label{bremcross}
\e^2{d\sigma\over d\e}={16\a r_0^2\e\over 3}{f_{nr}(x)\beta^{-2}+E\times
f_{\rm er}(x)\over 1+E}; \,\,\, x={\e\over E}. 
\eeq
Fig.~\ref{A1} compares this cross section (dashed-green) with the exact (3BN)
cross section of KM59 (solid-black). As evident the above simpler expression
agrees with the exact values very well with largest deviation of less than few
\% around energies $\e=mc^2$ and $E=mc^2$. This expression can be used for
analytic derivation of photon spectra. To include the contribution of
relativistic electron-electron bremsstrahlung  one should change $E\rightarrow
2E$ in the numerator.

\begin{figure}[!ht]
\begin{center}
\includegraphics[width=8.0cm]{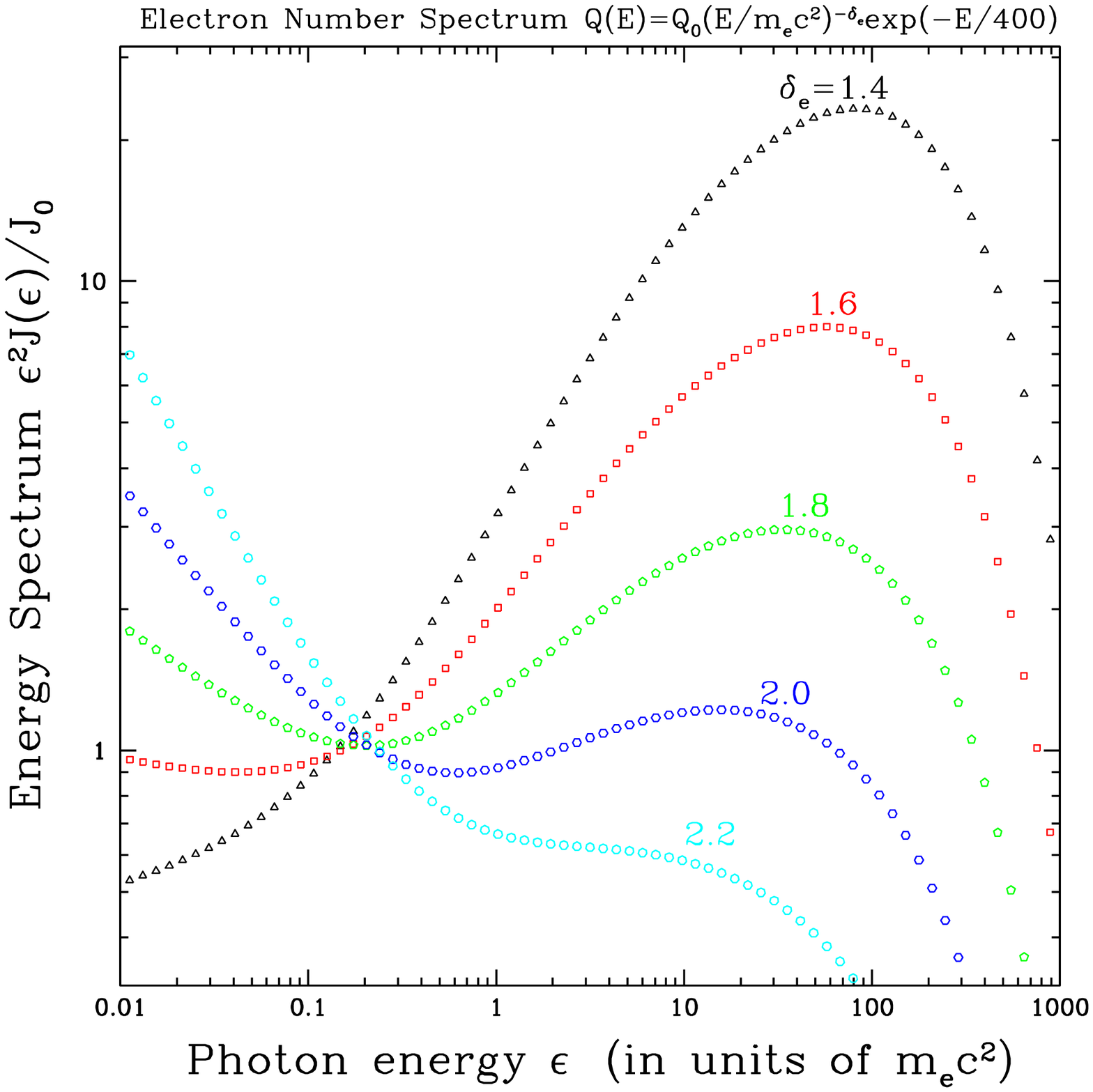}
\includegraphics[width=8.0cm]{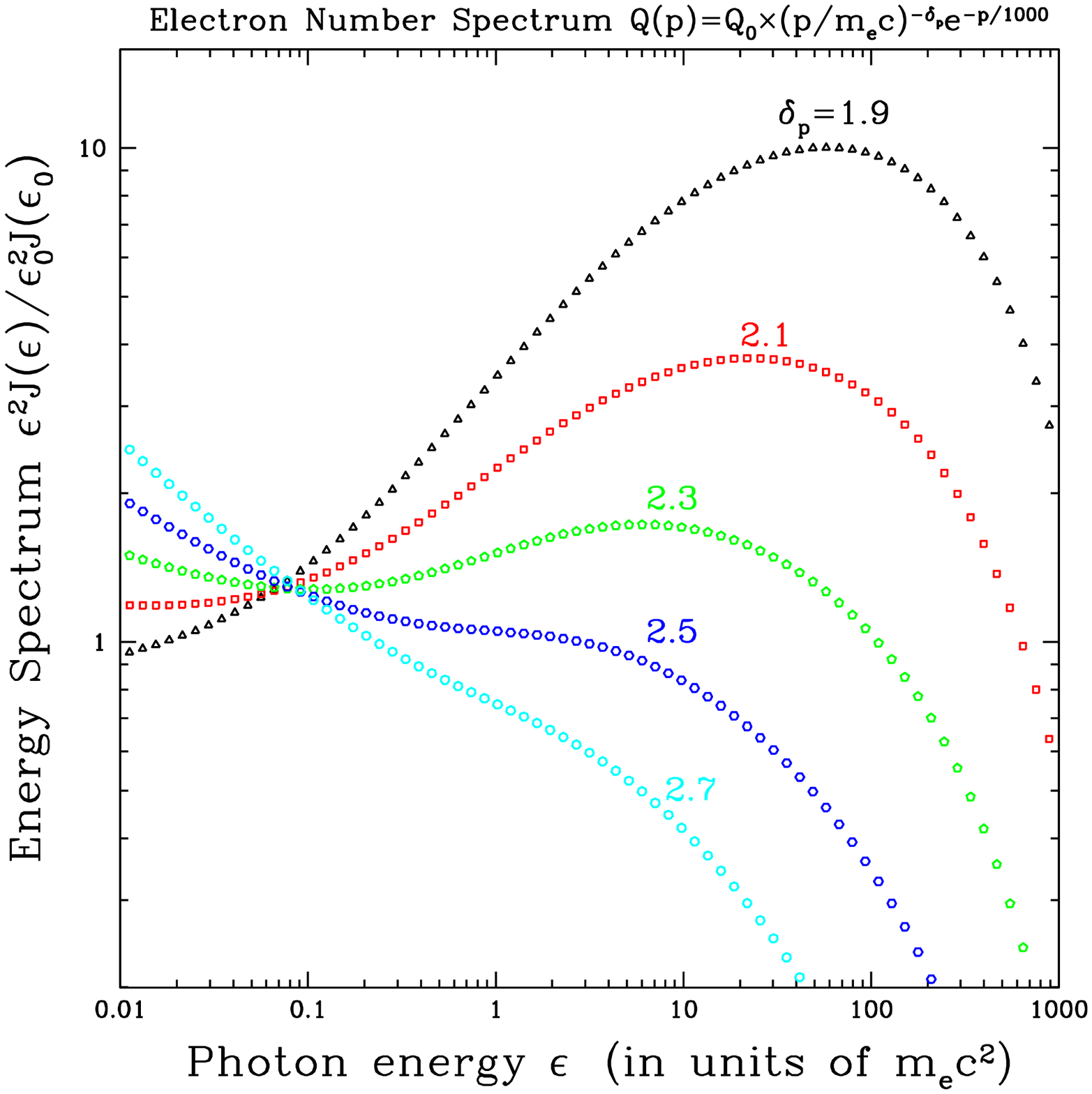}
\caption{Bremsstrahlung  energy spectra for power laws in energy (top) and
momentum (bottom) with exponential cut offs. For power laws in $E$  flat
spectra can be obtained  for $\d_e\sim 1.6$ only in the relativistic
regime and for $\d_e\sim 2.1$ only in the nonrelativistic regime. This due to
logarithmic dependence of $J(\e)\propto (\ln \e+ a)$ in the relativistic regime.
However, for power-law spectra in momentum this term is compensated by the
steeping of the spectra in the relativistic regime and fairly flat spectra can
be obtained in both regimes for $\d_p\sim 2.3$.}
\label{A2}
\end{center}
\end{figure}

2. {\it Flat $\nu f(\nu)$ Bremsstrahlung Spectra:} Fig.~\ref{A2} shows
$\e^2J(\e)\propto \int_\e^\infty\e^2(d\sigma/d\e)\beta(E)Q(E)dE$ NTB photon
spectra obtained for a power law (with
exponential cutoff) electron  spectra in energy and momentum
space. As evident flat photon spectra extending over several decades in photon
energy is not possible for such electron spectra in the energy space but can be
achieved for a power-law in momentum space for $\d\sim 2.3$. 

\section{Appendix B: Some Details of Synchrotron Emission}
\label{AB}

1. {\it Numerical coefficients:} In \S 3.2 we introduced two coefficients which
depend only on the spectral index of the electrons. In the relativistic regime
they are (see, e.g.~ Rybicki \& Lightman 1979) 
\beq\label{adelta}
a(\d)={3^{\d/2}\over \d+1}\G\left({3\d+19\over 12}\right)\G\left({3\d-1\over
12}\right)\langle\sin\t^{(\d+1)/2}\rangle
\eeq
and 
\beq\label{bdelta}
b(\d)={3^{(\d+1)/2}\over 2}\G\left({3\d+22\over 12}\right)\G\left({3\d+2\over
12}\right)\langle\sin\t^{(\d+2)/2}\rangle
\eeq
where $\G$ stands for the {\it Gamma} function and $\t$ is the angle between
the line of sight and the $B$ field. For the LT sources the magnetic field may
be radial or horizontal with respect to the limb  so that we have
$\t=\pi/2$ and the angular
terms are equal to one. In the opposite case of chaotic field lines 
the last terms in the above equations are equal to
$(\sqrt{\pi}/2)\G[(\d+5)/4))]/\G[(\d+7)/4]$ and
$(\sqrt{\pi}/2)\G[(\d+6)/4))/\G[(\d+8)/4]$, respectively. An accurate
determination of these coefficients is important
because the magnetic field estimates are sensitive to their values. Table 3
gives the values of these and other parameters for  the range $3<\d<5$ of
interest here.

\begin{deluxetable*}{ccccccc}[ht]
\label{tab:coeffs}
  \tablewidth{\textwidth}
  \tablecaption{Synchrotron Parameters}
  \tablehead{
  \colhead{Index $\d$} & \colhead{$a(\d)$} &  \colhead{$b(\d)$} &
\colhead{$c(\d)$} & \colhead{$\tau_p$} & \colhead{$f(\tau_p)$} &
\colhead{$cf(\tau_p)$}
}\\
\startdata
3 & 1.2 & 4.8  & 0.25 & 0.87(1.4) & 0.58(0.34) & 0.14(0.09)\\
4 & 2.0 & 6.6  & 0.30 &  0.95(2.0) & 0.61(0.43) & 0.18(0.13)\\
5 & 2.6 & 12  & 0.22 & 1.03(2.6) & 0.64(0.50) & 0.14(0.11)\\
\enddata
 \end{deluxetable*}
 
\vspace{0.3cm}

2. {\it Optical Depths and Magnetic Fields:} 

We are interested in the
spatially integrated flux 
\beq\label{fnu}
F(\nu)= S(\nu) \Omega f(\tau_\nu),
\eeq
where
$S(\nu)$ is the average source term and $f(\tau)$ depends on the shape and
geometry of the source. For example, for the plane-parallel approximation
$f(\tau)=1-e^{-\tau}$ and for a spherically symmetric source
$f(\tau)=1-2/\tau+2(1-e^{-\tau})/\tau^2$. Setting the derivative of the flux to
zero
we get $d\ln f(\tau)/d\ln \tau=5/(\d+4)$, and peak optical depths $\tau_p,
f(\tau_p)$
and $c(\d)f(\tau_p)$ shown in Table 3 for plane parallel
and spherical (in parenthesis) geometries. Inserting these values in
Eqs.~(\ref{fnu}) and using Eq.~(\ref{source}) we calculate gyrofrequency as
\beq\label{nub1}
\nu_b=\nu_p[c(\d)f(\tau_p)\D \Omega m\nu_p^2/F(\nu_p)]^2.
\eeq
Inserting the observed values shown in Tables 1 and 2 and the coefficients in
Table 3 we find gyro-frequencies and magnetic field values of 1.0(0.6) GHz and
360(220) G for Oct13, and 10(4) MHz and 3.6(1.5) G for Sep14 flares (spherical
geometry  in parenthesis). (Note that $c(\d)=a(\d)/b(\d)$, and hence the $B$
field, varies more slowly with the spectral index $\d$  [than $a(\d)$ and
$b(\d)$]
and the  angle $\t$ (it would change by 10\% going from random field to
ordered field with $\t=\pi/2$). 

We can obtain the  $B$ field with an {\it
alternative method} which is independent of $F(\nu_p)$, the  most uncertain
observationally determined parameter. In this method we first eliminate one of
the unknowns, namely $n_0$ using Eqs.~(\ref{opticaldepth}) and (\ref{thinF}) to
obtain $\nu_B$. From the first equation evaluated at $\nu_p$ and the second
equation at any frequency in the optically thin regime we get
\beq\label{n0L}
{\a n_0Lh\over 4\pi}={\tau_p m\nu_p\over
b(\d)}\left({\nu_p\over \nu_B}\right)^{1+\d/2}={F(\nu)\over
a(\d)\Omega\nu}\left({\nu\over \nu_B}\right)^{(1+\d)/2},
\eeq
which then gives
\beq\label{nub2}
\nu_B=\nu_p[c(\d)\tau_p\Omega m\nu^2/F(\nu)]^2(\nu_p/\nu)^{3+\d}. 
\eeq
Using the fluxes given in Table 1 and parameters in table 3  we get 
gyro-frequencies and magnetic field values of 1.6(9.7) GHz and
560(3500) G for Oct13, and 22(58) MHz and 7.9(21) G for Sep14 flares (spherical
geometry  in parenthesis). These values are sensitive to $\d$; e.g.~for $\d=5$
(instead of 4.7)
for Oct13 and $\d=3$ (instead of 2.7) for Sep14 we get $B=370(2300)$ and
$B=3.4(9.0)$,
respectively.  

\vspace{0.3cm}
3. {\it Free-free absorption:}

As mentioned above, free-free absorption with the absorption coefficient (see,
e.g.~Benz 1993)
\beq\label{ffkappa}
\kappa_{ff}=0.2 \nu^{-2} T^{-1.5}(EM/V)\left[1+0.05\ln{T/10^7 {\rm K}\over
\nu/{\rm GHz}}\right] 
\eeq
can be important for high densities and low magnetic fields.  For the Oct13
flare with large emission measure $EM\sim 1.3\times
10^{48}$ cm$^{-3}$ at $T=0.63\times 10^7$ K (obtained from \r data; Fatima
Rubio 2017, private
communication) we get an optical depth of $\tau_{ff}=50({\rm GHz}/\nu)^2$, where
we have used an area $A=V/L\sim 10^{18}$ cm$^2$, so that free free absorption
can be important below $\sim 7$ GHz. The dashed green curve in
Fig.~\ref{fig:radio}
shows a model spectrum that includes both free-free and synchrotron self-
absorption with a total optical depth
\beq\label{totkappa}
\tau(\nu)=(\nu/\nu_{ff})^{-2}+(\nu/\nu_{sy})^{-(2+\d)/2},
\eeq
so that $\tau_{ff}=1$
at $\nu_{ff}=7.5$ GHz and self
absorption, with  optical depth of unity at $\nu_{sy}=2.7$ GHz,  becomes 
dominant for
$\nu<1.2$ GHz for  electron index
$\d=5.2$. This will require $EM/(T^{1.5}A)\sim 3\times 10^{20}$ which is within
a factor of 3 of values quoted above. 
This  will change the
required magnetic field to a lower value. Following the steps of the second
(alternative) method used above, we can again eliminate $n_0$ and obtain $\nu_B$
and $B$. After some algebra we get  
\beq\label{ffB}
\nu_B=\nu_{ff}[c(\d)\Omega \nu_{sy}^2/F_0]^2=0.59 {\rm GHz}
\eeq
or $B=210$ G,  similar to the lower values obtained above. Since the inclusion
of free-free absorptions improves the fit to the data we will use this value of
the magnetic field. 

The above values of free-free absorption coefficients imply free-free
emissivity (in the microwave range) of $J_{ff}(\nu)=4\pi B(\nu,T)
\kappa_{ff}(\nu)$, where $B(\nu,T)=2kT(\nu/c)^2$ is the black-body brightness
in the Rayleigh-Jeans limit ($h\nu\ll kT$). It is easy to show  that the
expected free-free flux at optical depth of one (or
$\kappa(\nu_{ff}=(1-1/e)/L=0.6A/V$ is
\beq
\label{ff}
F(\nu>\nu_{ff})=2kT\Omega(1-1/e)(\nu_{ff}/c)^2\sim 5\, {\rm SFU},
\eeq
(for $T=0.6\times 10^7$ K, $nu_{ff}=7$ GHz and $\Omega \sim 10^{-8}$ sr) about a
factor of 6 below the observed synchrotron flux. 

In summary, averaging the above result we get magnetic field
values of 2 to 10 G (Sep14) and 200-500 G for Oct13 flares which we have
entered in the last column of Table 2.

It is interesting to note that this emission, extrapolated to
few keV range should also agree with the thermal HXR flux. This involves
extrapolation over a large frequency range (from $\sim 10^{10}$ to $10^{18}$ Hz)
and
differences between the Gaunt factors at microwaves of $\sim 15$ and HXRs
of unity. Nevertheless, dividing the above flux by this factor and the Boltzmann
factor $e^{\e/kT}\sim 10^{4.5}$ (for $\e=10$ keV and $T=10^7$ K), we get 10 keV
thermal  bremsstrahlung flux of $F(\e=10 keV)\sim \times 10^{-23}$ erg cm$^{-2}$
s$^{-1}$, Hz$^{-1}$ or $\nu f(\nu)$ flux of $10^{-3}$ vs the observed value of
$\sim 10^{-4}$
erg cm$^{-}2$ s$^{-1}$ (see, Fig.~3 in Pesce-Rollins 2015). Considering the
scale of
the extrapolation this is a satisfactory agreement.

\bibliographystyle{apj}
\bibstyle{aa}


\def\refer { \par \noindent \hangindent=2pc \hangafter=1}
\baselineskip = 10 true pt

\section*{REFERENCES}

\refer Ackermann  Ajello, M., Allafort, A., et al.\ 2012, \apj, 745, 144
\refer Ackermann, M., Ajello, M., Albert, A., et al.\ 2014 \apj, 787, 15
\refer Ackermann, M., Ajello, M., Albert, A,, et al.\ 2017 \apj, 835, 219
\refer Ajello, M., Albert, A., Allafort, A., et al.  2014 \apj, 789, 20
\refer Atwood, W.~B., Abdo, A.~A., Ackermann, M., et~al. 2009,  \apj, 697, 1071
\refer Barat, C., Trottet, G., Vilmer, N., et~al. 1994, \apjl, 425, L109
\refer Benz, A. D. 1993, {\it Plasma Astrophysics}, Kluwer Academic Publishers,
p 262
\refer Brown, J. C. 1972, \solphys, 25, 118
\refer Chen, Q., \& Petrosian, V. 2013, \apj, 777, 33
\refer Dulk, G.~A. 1985, \araa, 23, 169
\refer Effenberger, F., Rubio da Costa, F., Oka, M., et al. 2017, \apj, 835, 124
\refer Frost, K. J., \& Dennis, B. R. 1971, \apj, 165, 655
\refer Koch, H. W. \& Motz, J. W. 1959, Rev. Modern Phys., 31, 920
\refer Krucker, S., \& Lin, R.~P. 2008, \apj, 673, 1181
\refer Krucker, S., White, S.~M., \& Lin, R.~P. 2007, \apjl, 669, L49
\refer Krucker, S., Kontar, E. P., Christe, S., \& Lin, R. P. 2007, \apjl, 663,
L109 
\refer  Krucker, S., Hudson, H.~S., Glesener, L., White, S.~M., et al. 2010,
\apj, 714, 1108
\refer Lin, R.~P., \& Hudson, H.~S. 1971, \solphys, 17, 412
\refer Malyshkin, L., \& Kulsrud, R. 2001, \apj, 549, 402
\refer McTiernan, J.~M., \& Petrosian, V. 1990, \apj, 379, 381
\refer {Liu}, W., {Chen}, Q., \& {Petrosian}, V. 2013, \apj, 767, 168
\refer Maia, D. J. F., \&  Pick, M. 2004, \apj, 609, 1082
\refer  Masuda, S., Kosugi, T., Hara, H., Tsuneta, S., \& Ogawara, Y.\ 1994,
\nat, 371, 495
\refer Jin, M., Petrosian, V., Liu, W. et al. 2018, ApJ (in press); 
arXiv:1807.01427
\refer Nitta, N.~V., Freeland, S.~L., \& Liu, W.\ 2010, \apjl, 725, L28
\refer Petrosian, V. 1973, \apj, 186, 291
\refer Petrosian, V. 1981, \apj, 251, 727
\refer Petrosian, V. 1982, \apjl, 255, L85
\refer Petrosian, V. 2012, Space Sci Rev., 173, 535
\refer Petrosian, V. 2016, \apj, 830, 28
\refer Petrosian, V., \& Kang, B. 2015, \apj, 813, 5
\refer Petrosian, V., \& Donaghy, T.~Q. 1999, \apj, 527, 945
\refer Petrosian, V., Donaghy, T.~Q., \& McTiernan, J.~M.\ 2002, \apj, 569, 459
\refer Petrosian, V., \& Liu, S.\ 2004, \apj, 610, 550
\refer Pesce-Rollins, M.  Omodei, N., Petrosian, V., et al. 2015 \apjl, 805. 15
\refer Piana, M., Massone, A.~M., Kontar, E.~P., et~al. 2003, \apjl, 595, L127
\refer Plotnikov, I., Rouillard, A. P., \& Share, G. H. 2017, {\it A\&A}, 608, A43 
\refer Pryadko, J.~M., \& Petrosian, V.\ 1997, \apj, 482, 774 
\refer Ramaty, R. 1969, \apj, 158, 753
\refer Ramaty. R. \& Petrosian, V. 1972, 178, 241
\refer Rybicki, G.~B., \& Lightman\}, A.~P. 1979, {\it Radiative processes in
  astrophysics}
\refer Share, G. H.,  Murphy, R. J., Tolbert, A. K. et al.2017; arXiv:1711.01511
\refer Vestrand,  W.~T., \& Forrest, D.~J. 1993, \apjl, 409, L69
\refer Vilmer, N., Trottet, G., Barat, C., et~al. 1999, \aap, 342, 575

\end{document}